\UseRawInputEncoding

\documentclass[twocolumn]{aastex701}

\usepackage[utf8]{inputenc}
\usepackage{amsmath}
\usepackage{multirow}
\usepackage{graphicx} 
\usepackage{booktabs}
\usepackage{color}

\newcommand{\halpha}{H$\alpha$}
\newcommand{\hbeta}{H$\beta$}
\newcommand{\hgamma}{H$\gamma$}

\usepackage[T1]{fontenc}

\usepackage{scalerel}
\newcommand{\CaII}{{\ion{Ca}{2}}}

\newcommand{\CaIIk}{{\ion{Ca}{2}} K}

\newcommand{\MgII}{{\ion{Mg}{2}}}
\newcommand{\MgIIk}{{\ion{Mg}{2}} k}

\newcommand{\av}{A$_{\text{V}}$}
\newcommand{\mdot}{$\dot{\text{M}}$}
\newcommand{\Mdot}{{\dot{{M}}}}

\newcommand{\msun}{ M_{\sun}}

\newcommand{\msunyr}{M_{\sun} \, \rm{ yr^{-1}}}
\newcommand{\kms}{ \, km \, s^{-1}}

\newcommand{\ri}{R$_{\rm i}$}
\newcommand{\rw}{W$_{\rm r}$}

\newcommand{\tmax}{T$_{\rm max}$}

\received{2025, July 11}
\revised{2025, September, 5}
\accepted{2025, September, 11}
\submitjournal{ApJ}

\begin{document}

\title{
Vanishing Refractories: Tracing Dust Evolution in the BP Tau Protoplanetary Disk
}

\correspondingauthor{Marbely Micolta}

\author[0000-0001-8022-4378]{Marbely Micolta}
\affiliation{Department of Astronomy, University of Michigan, 1085 South University Avenue, Ann Arbor, MI 48109, USA.}
\email[show]{micoltam@umich.edu}

\author[0009-0006-7742-1761]{Justin Svarc}
\affiliation{Department of Astronomy, University of Michigan, 1085 South University Avenue, Ann Arbor, MI 48109, USA.}
\email{jsvarc@umich.edu}

\author[0000-0002-3950-5386]{Nuria Calvet}
\affiliation{Department of Astronomy, University of Michigan, 1085 South University Avenue, Ann Arbor, MI 48109, USA.}
\email{ncalvet@umich.edu}

\author[0000-0001-6647-862X]{Ezequiel Manzo-Martínez}
\affiliation{Instituto de Astronomía, Universidad Nacional Autónoma de México. Km. 107 Carretera Tijuana-Ensenada, Ensenada Baja California, México. C.P. 22860}
\affiliation{Department of Astronomy, University of Michigan, 1085 South University Avenue, Ann Arbor, MI 48109, USA.}
\email{emanzo@umich.edu}

\author[0000-0003-4507-1710]{Thanawuth Thanathibodee}
\affiliation{Department of Physics, Faculty of Science, Chulalongkorn University, 254 Phayathai Road, Pathumwan, Bangkok, 10330 Thailand.}
\email{Thanawuth.T@chula.ac.th}

\author[0000-0003-1166-5123]{Gladis Magris C.}
\affiliation{Centro de Investigaciones de Astronomía ``Francisco J. Duarte" CIDA, Av. Alberto Carnevali, Mérida 5101, Mérida, Venezuela.}
\email{gladismagris@gmail.com}

\author[0000-0003-4507-1710]{Avalon Littwiller}
\affiliation{
College of Engineering, University of Michigan, 1221 Beal Ave, Ann Arbor, MI 48109, USA.}
\email{litwild@umich.edu}

\begin{abstract}

We present a multi-wavelength analysis of the dust of BP Tau's protoplanetary disk. We use new optical spectra of BP Tau, taken with the Magellan/MIKE
spectrograph in tandem with archival UV and mid-infrared observations. We use the magnetospheric accretion model to analyze the \CaIIk\ and \MgII\ 2796.4 \AA\ emission lines and derive the abundance of Ca and Mg in the accretion flows as a proxy for the refractory abundance in the innermost gas disk. Furthermore, we used irradiated accretion disk models to compare the spectral energy distribution (SED) to observations and model in detail the 10$\mu$m and 20$\mu$m silicate features to obtain the spatial distribution and stoichiometry of the dust in which the refractories are locked in the disk. We find a significant degree of depletion of refractory material in the innermost gas disk with median abundances of $\rm [Ca/H] = -2.0^{+0.1}_{-0.0}$ and $\rm [Mg/H] = -1.30^{+0.2}_{-0.3}$ and attribute this to both radial drift and dust trapping due to a pressure bump/gap. Our SED modeling recovers the inner cavity that extends up to 8 AU, consistent with sub-mm observations.
We found a significant decrease of the Mg-to-Fe ratio with decreasing radius, with Mg-rich silicates in the outer wall and Fayalite in the inner wall, consistent with the Mg depletion inferred from the emission lines.

\end{abstract}


\section{Introduction}

Throughout the evolution of protoplanetary disks, the initially gas-rich disks 
lose mass by being accreted onto the central star, while simultaneously losing material through winds and photoevaporation \citep[e.g.,][]{hartmann_accretion_2016}. Simultaneously, solid particles within the disk accumulate, grow, settle, and migrate inward, eventually forming planetesimals and planets \citep{weidenschilling_origin_1997,brauer_coagulation_2008,birnstiel_simple_2012}.

Because refractory (rocky) material requires the highest temperatures to sublimate, they remain locked in solid particles of mm/cm size (pebbles) until they reach the dust-sublimation front (dust wall) at $\sim$ 0.1 au. Therefore, a snapshot of the distribution of the rocky material throughout the disk provides a crucial record of the stage of dust evolution, and the effects of radial transport and trapping in the disk
\citep[e.g., radial drift, pressure bumps,][]{drazkowska_planet_2023},
key mechanisms for setting the conditions for planet formation to begin in the system. Radial drift will create a brief period of refractory enrichment in the inner region of the disk, followed by depletion as the larger grains containing the rocky material reach the inner disk in fast timescales and are accreted onto the star \citep{huhn_how_2023}. Pressure bumps block the inward drift of pebbles, further depleting the inner regions of refractory elements. This effect has been observed in Herbig Ae/Be stars \citep[][]{kama_fingerprints_2015, guzman-diaz_relation_2023}, T Tauri Stars \citep[TTSs;][]{mcclure_carbon_2019,mcclure_measuring_2020, micolta_ca_2023,micolta_using_2024}, and Brown Dwarfs \citep{france_metal_2010}. 

Once ongoing, planets in the early stages of formation leave their marks on the rocky material of the disk as they sequester pebbles to build their cores
\citep[e.g.,][]{lambrechts_rapid_2012, drazkowska_planet_2023}, 
and block inward drift \citep[e.g.][]{pinilla_trapping_2012, zhu_dust_2012, van_der_marel_diversity_2021},
reducing the abundance of rocky material that will reach the inner regions of the disk \citep[][]{schneider_how_2021,schneider_how_2021-1,huhn_how_2023}. To unravel the possible causes for the depletion of rocky material in the inner regions, we need to measure refractory depletion in inner disks while also looking for signs of radial drift and pressure-bump effects on the disk.

In this work, we take BP Tau as a template for future studies.
BP Tau is a well-characterized classical T Tauri star, with a compact disk initially thought to have no structure \citep{long_compact_2019}, but now known to have a gap inside 10 au, revealed by
higher resolution analysis of ALMA data and new observations
\citep{zhang_substructures_2023,yamaguchi_alma_2024,gasman_minds_2025}.
We follow \citet{micolta_ca_2023,micolta_using_2024} and obtain the abundance of refractory material reaching the star from the analysis of the \MgII\ and \CaII\ emission lines formed at the accretion flows. We will perform a detailed analysis of the spectral energy distribution (SED) following \citet{mcclure_curved_2013} and \citet{mauco_herschel_2018} to inform us of the structure, radial gradients of dust stoichiometry, and grain sizes. In particular, the stoichiometry of the silicate features will provide a direct comparison between the abundance of Mg through the disk and the abundance observed at the accretion flows. 

This paper is organized as follows. In Section \ref{sec:data} we describe the observations taken and the data used. In Sections \ref{sec:magneto-mod} and \ref{sec:diad} we describe the magnetospheric accretion model and the disk model, respectively.  In Section \ref{sec:results}, we present our analysis and results. In Section \ref{sec:dis}, we discuss the implications of our results. Finally, in Section \ref{sec:con}, we give our conclusions.

\section{Data}\label{sec:data}

\subsection{New Observations} 

We obtained an optical spectrum of BP Tau on November 26, 2021 with the Magellan Inamori Kyocera Echelle \citep[MIKE,][]{bernstein_mike_2003} spectrograph on the Magellan Clay telescope at the Las Campanas Observatory in Chile. We used the 0.7" slit and the blue and red cameras with a spectral resolution of R $\sim$ 32,500 for a wavelength range of 3200 to 10000 {\AA}. We used an exposure time of 1200s for each camera, obtaining a median signal-to-noise ratio of the line and continuum of 33 and 237 between 3928 - 3938{\AA} and 6550 - 6574 {\AA}, respectively. We reduced the data using the CarPy package \citep{kelson_evolution_2000,kelson_optimal_2003}.

\subsection{Supplementary Data} 

\subsubsection{Ultra-violet spectrum} 

BP Tau was observed with the Goddard High Resolution Spectrograph (GHRS) instrument on the Hubble Space Telescope (HST) on 30 July 1993 as part of the GO 3845 program (PI: Basri, Gibor), using the G270M grating, obtaining a median resolution of R $\sim$ 20,000 from 1150 to 3200 {\AA}. This allowed us to resolve the Magnesium \ion{Mg}{2} doublet at 2800 {\AA}, which was not possible with more recent observations obtained at lower resolution. We downloaded the spectrum from the Mikulski Archive for Space Telescopes (MAST)\footnote{spectra available in MAST dataset:\dataset[10.17909/hma8-8838]{http://dx.doi.org/10.17909/hma8-8838} \label{fnote:1}}. These data had been reduced and calibrated by the pipeline of their instruments.

\begin{deluxetable*}{lcccc}[t!]
\label{tab:photometry}
\tablecaption{Photometry}
\tablehead{
\colhead{Filter} & 
\colhead{$\lambda$ ($\mu$m)} & 
\colhead{$\lambda F_{\lambda}$} & 
\colhead{$\lambda F_{err, \lambda}$ } & 
\colhead{Reference}
} 
\startdata
2MASS:J & 1.24 & 1.2e-9 & 3.9e-11 & \citet{zacharias_naval_2004} \\
2MASS:H & 1.65 & 1.2e-09 & 2.7e-11 &  \citet{zacharias_naval_2004}\\
2MASS:Ks & 2.16 & 8.6e-10 & 1.9e-11&  \citet{kharchenko_all-sky_2001} \\
WISE:W1 & 3.35 & 4.1e-10 & 1.0e-12 &  \citet{cutri_vizier_2021}\\
Spitzer:IRACI2 & 4.49 & 2.1e-10 & 1.3e-12 & \citet{esplin_survey_2019}\\
Herschel:PACS70 & 70.0 & 2.6e-11 & 4.3e-12 &  \citet{ribas_far-infrared_2017}\\
Herschel:PACS160 & 160.0 & 9.4e-12 & 1.9e-12 &  \citet{ribas_far-infrared_2017}\\
Herschel:SPIRE250 & 250.0 & 5.8e-12 & 1.2e-12&  \citet{ribas_far-infrared_2017} \\
Herschel:SPIRE350 & 363.0 & 3.3e-12 & 6.6e-13&  \citet{ribas_far-infrared_2017} \\
Herschel:SPIRE500 & 517.0 & 1.7e-12 & 3.5e-13&  \citet{ribas_far-infrared_2017} \\
SCUBA & 849.0 & 4.6e-13 & 2.5e-14 & \citet{mohanty_protoplanetary_2013}\\
SMA & 887.0 & 4.4e-13 & 4.1e-14 & \citet{andrews_mass_2013}  \\
SCUBA & 1300.0 & 1.1e-13 & 1.7e-15& \citet{mohanty_protoplanetary_2013} \\
SMA & 1330.0 & 9.4e-14 & 4.9e-15 & \citet{andrews_mass_2013} \\
VLA & 40000.0 & 1.1e-17 & 1.2e-18&   \citet{dzib_goulds_2015} \\
VLA & 66600.0 & 4.1e-18 & 9.9e-19&   \citet{dzib_goulds_2015} \\
\enddata
\tablecomments{$\lambda F_{\lambda}$ in units of $\rm erg\ cm^{-2}\ s^{-1}$}
\end{deluxetable*}

\subsubsection{Mid-Infrared spectrum} 

\label{sec:midir}

BP Tau was observed with the Spitzer InfraRed Spectrograph \citep[IRS,][]{houck_infrared_2004, werner_spitzer_2004} in 2008 as part of Program 50537
(PI:Boden).
It was also observed in 2023
with
the JWST Mid-Infrared Instrument \citep[MIRI,][]{rieke_mid-infrared_2015,wright_mid-infrared_2015,wright_mid-infrared_2023}, as part of the MIRI Mid-INfrared Disk Survey \citep[MINDS; PID:1282,][]{henning_minds_2024}. 
We downloaded the IRS spectrum, AORkey: 26354432, from the Combined Atlas of Sources with Spitzer IRS Spectra (CASSIS) database \citep{lebouteiller_cassis_2011}.
The JWST-MIRI spectrum\textsuperscript{\ref{fnote:1}} reduction process is explained in \citet{gasman_minds_2025,temmink_minds_2025}.

\begin{figure}[t!]
\epsscale{1.19}
\plotone{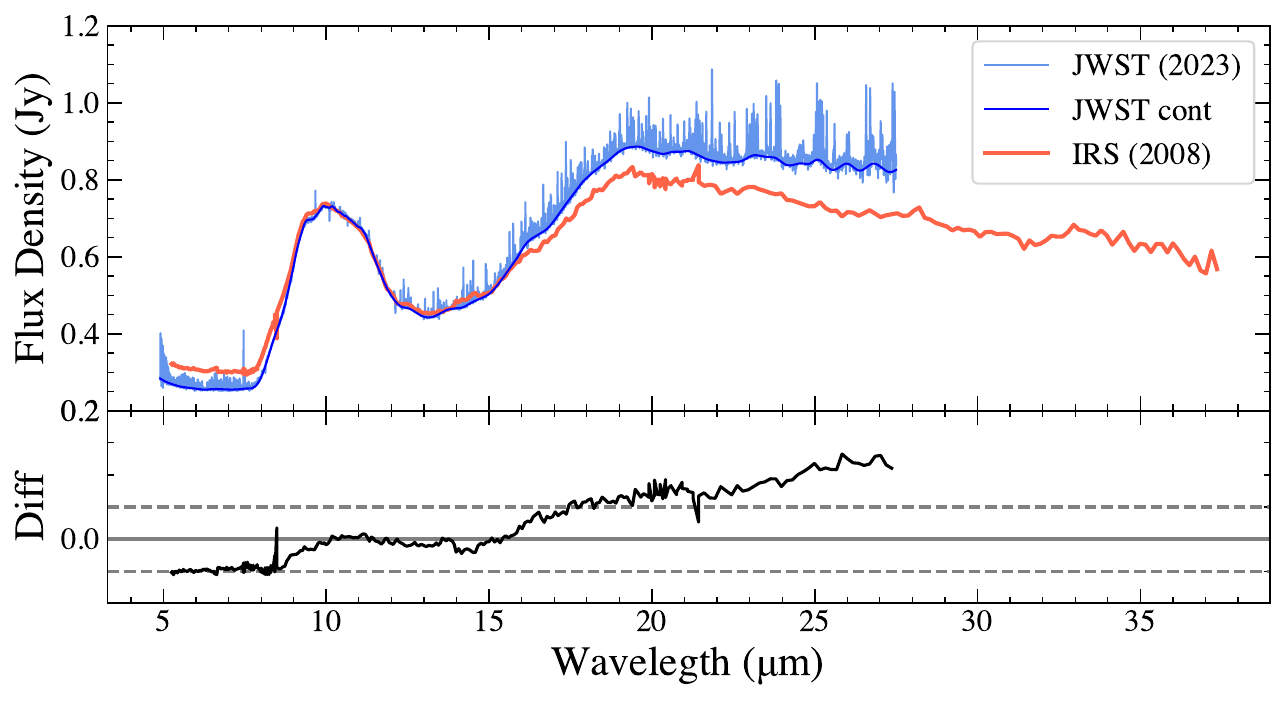}
\caption{Comparison between JWST-MIRI and Spitzer-IRS spectra of BP Tau. Top: Red line shows the Spitzer-IRS spectra (R $\sim 60-100$).
Dashed blue lines show the full wavelength coverage of the JWST-MIRI at its full resolving power(R$\sim 1600-3400$), solid blue line shows JWST-MIRI spectra convolved to the IRS resolution. Bottom: Difference between IRS and the continuum of the JWST-MIRI spectra (black solid line). Grey solid line marks a difference equal to 0, dashed grey lines mark difference of $\pm 0.05$ ($5\%$). 
}
\label{fig:jwst-vs-irs}
\end{figure}

Figure \ref{fig:jwst-vs-irs} shows a comparison between the JWST-MIRI (blue) and Spitzer-IRS BP Tau spectra (red), while the bottom panel shows the difference between the IRS and the JWST-MIRI spectra continuum.
Dashed lines represent a $5\%$ difference between the spectra.
The spectrophotometric accuracy for Spitzer IRS is $2-10\%$ \citep{furlan_survey_2006,watson_crystalline_2009}, while for JWST/MIRI MRS it is $5.6\% \pm 0.7\% $ \citep{argyriou_jwst_2023}.  
Overall, the JWST spectrum shows a decrease in flux at shorter wavelengths ($<8\mu m$), but an increase in flux at longer wavelengths ($>16\mu m$), with the $10\mu m$ silicate feature remaining unchanged. This is reminiscent of seesaw continuum variability, in which an increase in flux at longer wavelengths corresponds to a decrease at shorter wavelengths and vice versa. This can be explained by changes in the height of the inner wall of the disk \citep[][]{muzerolle_evidence_2009,espaillat_spitzer_2011, flaherty_highly_2011} and/or a misalignment between the inner and outer disks \citep[][]{espaillat_evidence_2024}, resulting in differences in the shadow cast on the edge of the outer wall. We further discuss the possibility of a misalignment in BP Tau \S \ref{sec:misalignment}.

Given that the science goal of this work is to analyze the dust composition throughout the inner disk, the fact that there are no significant changes in the silicate features, and the fact that the IRS spectra include the $30-38\mu m$ wavelength range, key to constraining the crystalline-silicate composition of the system (see \S\ref{sec:sedmodel}, Fig. \ref{fig:crystals}), our modeling will focus only on the IRS spectra.

\subsubsection{Photometry} 

We collected the photometry using the VizieR catalog access tool \citep{ochsenbein_vizier_1996}. We include the Two Micron All Sky Survey (2MASS), the Wide-field Infrared Survey Explorer (WISE), Spitzer/IRAC, Herschel/PACS, and Herschel/SPIRE. We also include the millimeter photometry available from the Submillimeter Common-User Bolometer Array on the James Clerk Maxwell Telescope \citep[SCUBA;][]{holland_scuba-2_2013} and the Submillimeter Array \citep[SMA;][]{ho_submillimeter_2004}. See Table \ref{tab:photometry}.

\section{Magnetospheric accretion models}
\label{sec:magneto-mod}

We applied {\it CV-multi} \citep{muzerolle_emissionline_2001} to determine the structure and emission of the magnetospheric accretion flows. In this framework, the accretion flows follow the stellar magnetic field lines, with a dipolar geometry characterized by the inner radius (\ri) and the width at the base of the flow \added{(\rw) }in the disk.  The density distribution is determined by the geometry and the mass accretion rate (\mdot). The temperature distribution is semi-empirical, calculated by balancing a volumetric heating rate ($r^{-3}$) and an optically thin cooling law \citep[][]{hartmann_magnetospheric_1994}; parameterized
by the maximum flow temperature (\tmax), which is constrained by the accretion rate, following the prescription of \citet[][]{muzerolle_emissionline_2001}. 

The models use the extended Sobolev approximation to calculate mean intensities, which in turn are used to calculate the radiative rates in the statistical equilibrium equations to obtain the level populations. Line profile calculations are performed using the ray-by-ray method for a given inclination angle $i$ and assuming Voigt profiles \citep{muzerolle_emissionline_2001}, with damping constants from \citep{vernazza_structure_1976}.

\subsection{{\halpha} and {\CaII} Implementation}

The physics and implementation of the magnetospheric accretion models for hydrogen and calcium are given in \citet{hartmann_magnetospheric_1994,muzerolle_emissionline_2001,micolta_ca_2023,micolta_using_2024}. Here, we highlight the basic assumptions: We adopt a 16-level hydrogen atom and a 5-level calcium II atom to obtain level populations, optical depths, and source functions.

\subsection{{\MgII} Implementation}

We calculate the population of Mg II by solving the statistical equilibrium equation for three ionization stages, assuming that all electrons are on the ground excitation level in each stage. To calculate the radiative
ionization rates, we used photoionization cross-sections from \added{\citet{verner_analytic_1995}}, and an ionizing radiation field covering from the X-ray to the FUV.
For the FUV (1100 - 2000 {\AA}), we used the spectrum of TW Hya, including the reconstructed Ly $\alpha$ \citep{herczeg_farultraviolet_2002}, scaled to the adopted accretion luminosity of BP Tau. For the X-rays, we took the observed SED of TW Hya \citep{kastner_evidence_2002},
scaled to a luminosity $L_X = 10^{30} {\rm erg \, s^{-1}}$
consistent with the range of 
$L_X$ observed for BP Tau,
0.7 $\times 10^{30}$ \citep{neuhaeuser_t_1995} to
4 $\times 10^{30}$ $\rm erg \, s^{-1}$ \citep{walter_smothered_1981}.
To simulate the unknown EUV fluxes,
we joined the FUV and LX spectra by linearly interpolating between the adopted FUV and the X-ray luminosities in log scale.
We attenuated the ionizing radiation field at each point in the magnetosphere, 
calculating the radial column density to the closest inner boundary of the magnetosphere and using this to calculate the optical depth for H bound-free and free-free transitions, the most important sources of continuum opacity in the magnetosphere.
Collisional ionization rates were taken from \citet{vernazza_structure_1976}.  
We represent the Mg II ion by a 2-level atom, and solve the populations, source function, and radiation fields using the extended Sobolev approximation \citep{hartmann_magnetospheric_1994}.

\subsection{Model grid}

\begin{deluxetable*}{lcc}[t!]
\label{tab:bptau_param}
\tablecaption{Stellar and Disk Properties of BP Tau}
\tablehead{\colhead{Property} & \colhead{Value} & \colhead{Reference} 
}
\startdata
Spectral type (SpT) & K7 & \citet{ingleby_accretion_2013} \\
Distance (d) & 129.1 pc & \citet{gaia_collaboration_gaia_2018} \\
Radius (R) & $\rm 2.1 R_{\odot}$ & \citet{ingleby_accretion_2013} \\
Luminosity (L) & $\rm 1.0 L_{\odot}$ & \citet{ingleby_accretion_2013} \\
Extinction (\av) & $1.1$ & \citet{ingleby_accretion_2013} \\
Mass (M)& $\rm 0.8 M_{\odot}$ & \citet{ingleby_accretion_2013} \\
Inclination ($i$) & $38^{\circ}$ & \citet{long_compact_2019} \\
Accretion rate (\mdot) & $\rm 2.9\times10^{-8} M_{\odot}yr^{-1}$ & \citet{ingleby_accretion_2013} \\
\enddata
\end{deluxetable*}

We calculated a large grid of magnetospheric models using BP Tau stellar parameters (Table \ref{tab:bptau_param}). The full parameter space explored is described in Table \ref{tab:model_param}; for each combination of magnetospheric parameters, we calculated the line profiles for H$\alpha$, \CaII\ lines, and \MgII\ lines. The abundance of Ca and Mg relative to H in the flows is left as a free parameter, and we cite it relative to the solar value \citep[$\rm log \left(N_{i} / N_{H}\right)_{\odot} = 6.34 \pm 0.04$ and $7.60 \pm 0.04$ \added{for $N_{i}$ being} Ca and Mg, respectively;][]{asplund_chemical_2009}. Line profiles were calculated for thirteen values of Ca and Mg abundances by number: 1 (solar abundance), 0.75, 0.5, 0.25, 0.1, 0.075, 0.050, 0.025, 0.01, 0.0075, 0.0050, 0.0025, 0.001 (0.1\% of solar abundance).
The grid has a total of 734832 profiles.

\begin{deluxetable}{lccc}[t!]
\tablecaption{Parameter space of magnetospheric accretion models  \label{tab:model_param}}
\tablehead{
\colhead{Parameters} & \colhead{Min.} & \colhead{Max.} & \colhead{Step} 
}
\startdata
log {\mdot} ($\msunyr$)	    & -7.00	& -8.25 & 0.25  \\
T$_{\rm max}$ (K)	        & 7500 & 10000	& 500 \\
R$_{\rm i}$	(R$_{\star}$)	& 1.50	& 3.50	& 0.25 \\
\added{\rw }(R$_{\star}$)	& 0.50   & 1.75   & 0.25 \\
$i$	(deg)				    & 34	& 78	& 4 \\ 
$\rm [X/H]$	& 0	&  -3	& ... \\ 
\enddata
\tablenotetext{*}{$\rm [X/H] = log (N_{X}/N_{H}) - log (N_X/N_{H})_{\sun}$, where X can be Ca or Mg}

\end{deluxetable}

\section{Disk models}\label{sec:diad}

To complement the determination of refractory abundances of material falling onto the star,  we modeled the emission of the disk, aiming to determine the spatial distribution, morphology, stoichiometry, and species abundances of the solids in which the refractories were locked. 
We adopted the stellar and accretion data from \citep{ingleby_accretion_2013}, shown in Table \ref{tab:bptau_param}.

\subsection{Disk structure}
We used the D'Alessio Irradiated Accretion Disks (DIAD) model \citep{dalessio_accretion_1998,dalessio_effects_2006} to fit the spectral energy distribution (SED) of BP Tau. This model assumes that the disk is steady and is heated by viscous dissipation and stellar and accretion shock irradiation. With these heating agents, the radial and vertical density and temperature structures are calculated self-consistently from hydrostatic equilibrium. We assume that the gas and dust temperature are equal
and solve the equations of energy conservation and transport following \citet{dalessio_effects_2006}. We assume two populations of dust grains, each with a size 
distribution given  by $n(a) \propto a^{-3.5}$ between a minimum size $a_{min} = 0.005 \mu$m and a maximum size $a_{max}$, which is a
free parameter. The populations have different vertical distributions; the large grains concentrate on the midplane, while small grains dominate in the upper layers. 

The dust \added{settling} distribution in the disk is characterized by the parameter $\epsilon$ that gives the dust-to-gas mass ratio of the small grains relative to the total dust-to-gas mass ratio, which depends on the abundances adopted for the species included \citep{dalessio_effects_2006}. We assume that at a given radius, the small dust mass that disappears from the upper layers is added to the large grains, so the total dust-to-gas mass ratio is constant \citep{dalessio_effects_2006}.
Moreover, we assume a mixture of silicates, carbonaceous materials and H$_2$O ice, and  condensation temperatures of 1400K for silicates, 425K for organics, and 180K for H$_2$O ice \citep{dalessio_effects_2006,pollack_composition_1994}.

\subsection{Dust opacities \label{sec:optool}}

We include amorphous and crystalline silicates; specifically, we include olivine (${\rm Mg}_{2x} \, {\rm Fe}_{2(1-x)} {\rm SiO_4}$) and pyroxene
(${\rm Mg}_x \, {\rm Fe}_{1-x} {\rm SiO_3}$),
and their crystalline forms
forsterite and enstatite, respectively, 
with
$x$~=~${\rm Mg / (Mg + Fe)}$, the magnesium content.

We calculate opacities using OPtool \citep{dominik_optool_2021}\footnote{https://ascl.net/2104.010}, which allows different methods to obtain absorption coefficients, and includes indices of refraction of many relevant materials, taking into account
different stoichiometries. We
use optical constants from OPtool and calculate mass absorption coefficients with the distribution of hollow spheres (DHS) method and the default volume fraction of 0.8 for amorphous materials and continuous distribution of ellipsoids (CDE) for crystalline materials.
For olivine, the available Mg fractions are 40 and 50\%,
while for pyroxene, the available Mg fractions are 40, 50, 60, 70, 80, 70, 80, 95, and 100\%.
For forsterite, the Mg fractions are 95, 100, and 0\% (which is formally fayalite). Only one Mg abundance, 96\%, is available for enstatite.
For organics, we calculate the mass absorption coefficients with the OPtool CHON amorphous material. We also include silica (SiO$_2$) with opacities from \citet{sargent_dust_2009} and water ice with opacities from \citet{dalessio_accretion_2001}.

\section{Analysis and Results} \label{sec:results}

\subsection{The Reduced $\chi^2$ Method \label{sec:chi2}}

For the analysis of both emission lines and SED, we use the reduced minimum $\chi^2$ ($\chi^2_{\nu}$ ) to evaluate the fit of a model $F(x)$ to the observed data $y$, following the equation:

\begin{equation}
\rm \chi^2_{\nu} = \frac{1}{n} \ \sum_{i=1}^n\left[\frac{\left(y_i-F_i(x)\right)^2}{\sigma_i^2} \right]
\label{eq:chi2}
\end{equation}
where $n$ is the number of points and $\sigma_i$ represents the uncertainty of the observations. In the following subsections, we discuss the specific details for the modeling of each case and the results obtained.

\subsection{Emission lines modeling}\label{sec:line-mod}

We used the magnetospheric accretion model grid (described in \S \ref{sec:magneto-mod}) to perform a detailed line profile modeling of the \halpha, \CaIIk\ and \MgII\ 2796.4 \AA\ lines. Since MIKE observations are not flux calibrated, the \halpha\ and \CaIIk\ lines are fitted using the flux normalized to the continuum. In the case of \MgII\ 2796.4 \AA, the profile is fit using the line luminosity. Therefore, each line is modeled individually.

Due to small observational errors, we introduce an error inflation parameter $b$ to account for the possible underestimation of uncertainties \citep[e.g.][]{hogg_data_2010,line_uniform_2015}. For each data point $i$: $s_i^2=\sigma_i^2+10^b$, where $\sigma_i$ represents the original uncertainty of the observations, \added{and $b$ is the error inflation parameter.} In order to fit the lines while also fitting the inflation parameters, we fit the line profiles using a maximum likelihood approach including the normalization factor in addition to the $\chi^2$ term, to prevent $b$ from approaching infinity. See Appendix \ref{ap:fitting-lines} for a more detailed discussion.  
\added{The best-fit parameters along with those obtained from the top-25 best-fitting models distribution for each line are in Table \ref{tab:lines_results}, along with the reduced $\chi^2$ obtained for each case. These models provide the best agreement with the observed data. In the next section, we discuss the modeling results for each line individually.}

\subsubsection{\halpha\ line} 

\begin{figure*}[t!]
\epsscale{0.98}
\plotone{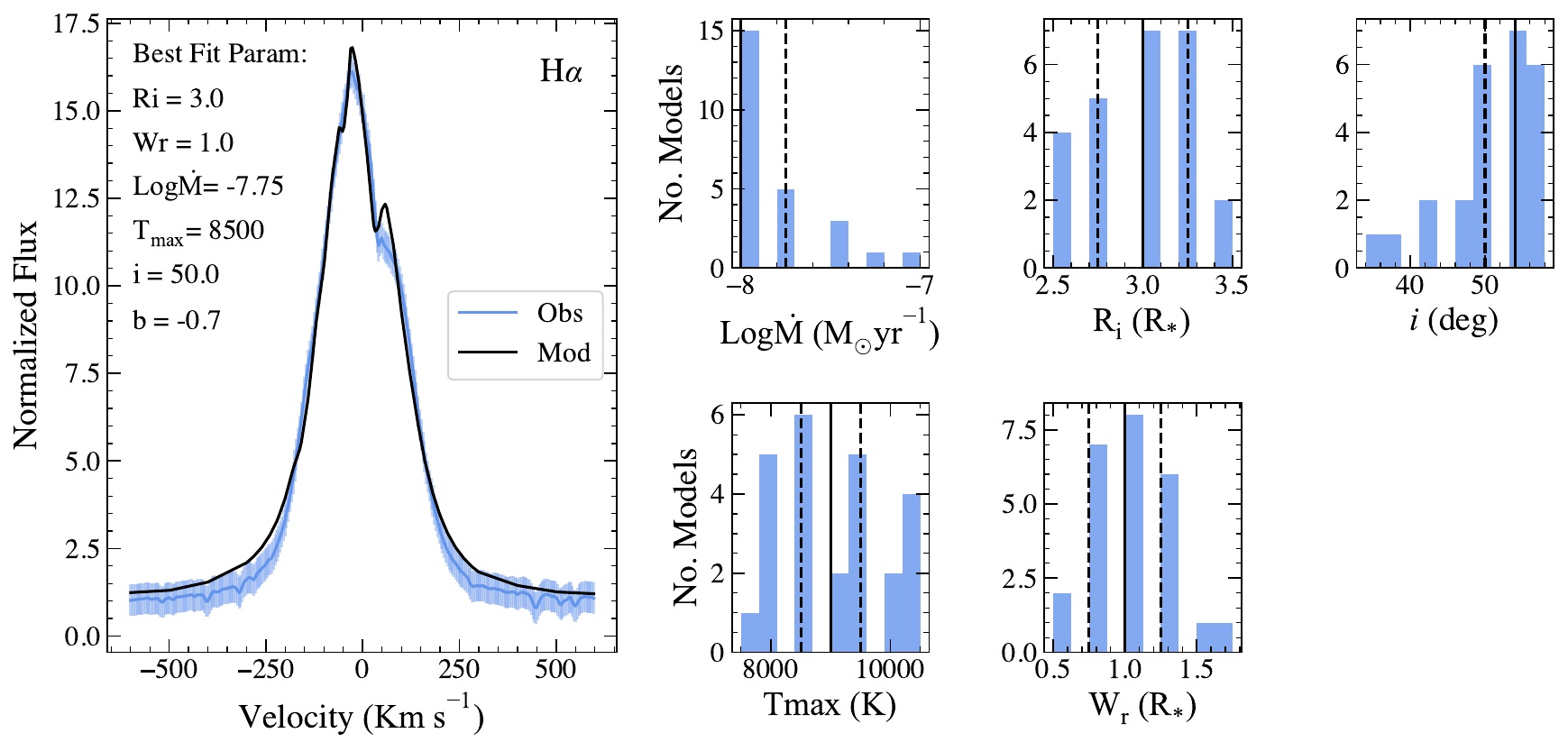}
\caption{Left: Comparison of the best-fit model and the observations for \halpha. Right: Parameter distribution for the top 25 best-fitting models. The solid line is the median value for each parameter, dashed lines represent the 25th and 75th percentile. }
\label{fig:halpha}
\end{figure*}

Figure \ref{fig:halpha} shows the profile comparison between the observations and the best-fit model (left) and the parameter distributions for the top 25 best-fitting models (right-hand histograms). The best-fit model has a $\rm \log \Mdot = -7.75\ \msunyr$ \added{(cf. Table \ref{tab:lines_results}.)}, which is consistent, within the uncertainties, with the one reported in 
\citet[][$\rm \log \Mdot = -7.54\ \msunyr$, see Table \ref{tab:bptau_param}]{ingleby_accretion_2013} and the median values measured in the 2021/2022 epochs of the TTSs ODYSSEUS survey monitoring campaign, with shock modeling predicting $\rm \log \Mdot = -7.88/-7.85 \ \msunyr$ and \halpha\ profile modeling $\rm \log \Mdot = -7.69/-7.65 \ \msunyr$ \citep[][, respectively]{wendeborn_multiwavelength_2024,wendeborn_multiwavelength_2024-1}.

Our modeling yields the \halpha\ 
line
originating from extended geometries, \ri $= 3.00 R_*$ and \rw$= 1.00 R_*$, both values are within the range obtained from the \halpha\ modeling in the monitoring campaign \citep[\ri$ = 2.7-4.1 R_*$, and \rw$ = 0.3-1.5 R_*$,][]{wendeborn_multiwavelength_2024-1}. The inner radius obtained is also consistent with the truncation radius obtained from 3D magnetospheric accretion models \citep[$3.6 -7 R_*$, highly dependent on the stellar parameters, especially \mdot,][]{long_global_2011}. In addition, both in this work and in \citet{wendeborn_multiwavelength_2024-1} 
it is found that the \halpha\ profile is only reproducible with inclinations of $50^\circ$ and $45^\circ$, respectively, which are higher than the known inclination of the BP Tau disk ($38^\circ$, Table \ref{tab:bptau_param}).

\subsubsection{\CaIIk\ line}

\begin{figure*}[t!]
\epsscale{0.98}
\plotone{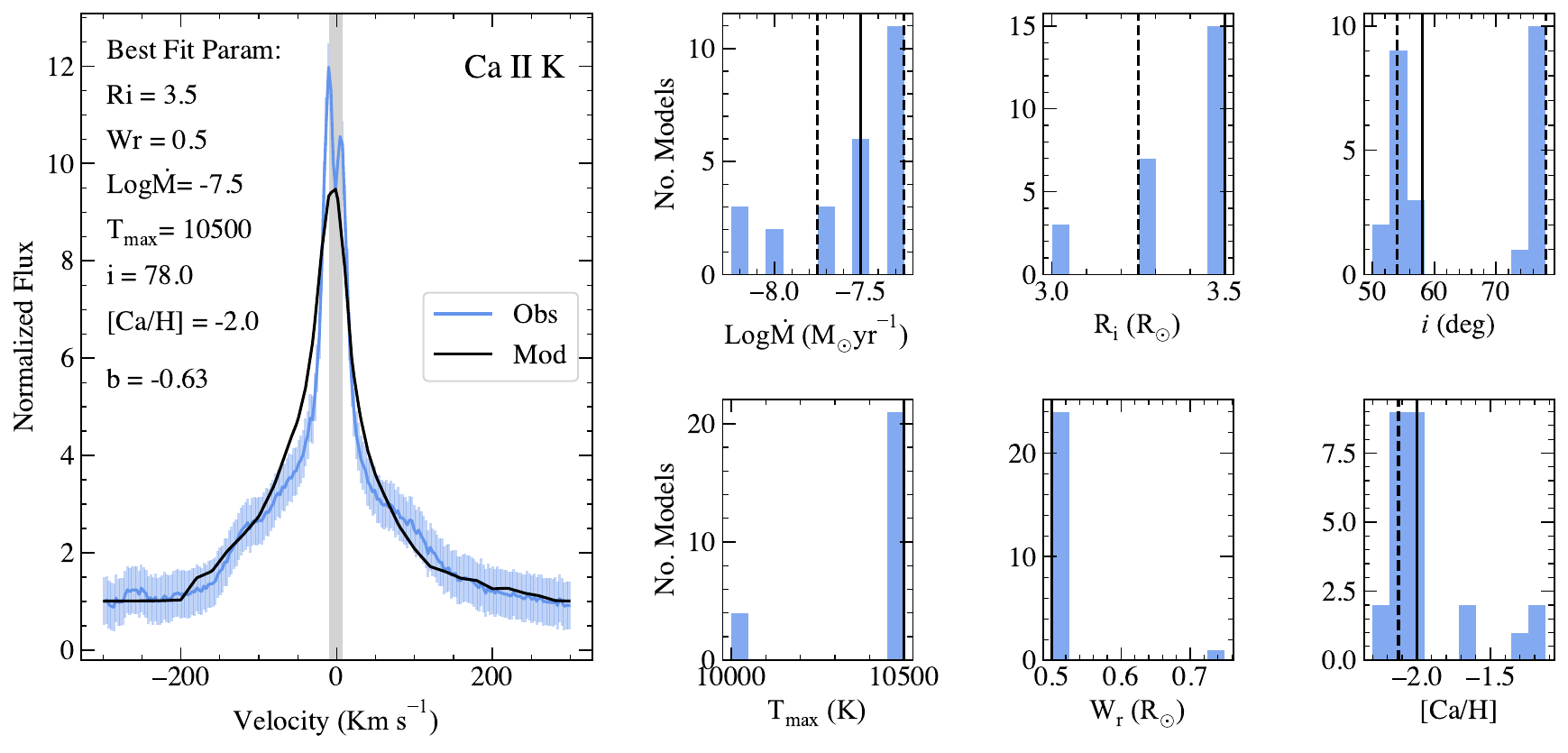}
\caption{Left: Comparison of the best-fit model and the observations for \CaIIk. Grey regions are excluded from the fitting. Right: Parameter distribution for the top 25 best-fitting models. The solid line is the median value for each parameter, dashed lines represent the 25th and 75th percentile. }
\label{fig:cak-re}
\end{figure*}

To perform the profile fitting for \CaIIk, we exclude the region $-8$ to $+7$ $\rm \kms $ where the central absorption feature is located. \added{This feature is consistent with it being formed in the interstellar medium \citep{draine_physics_2011}, which the model does not account for.}
Figure \ref{fig:cak-re} shows the profile comparison between the observations and the best-fit model (left) and the parameter distributions for the top 25 best-fitting models (right-hand histograms).\added{We detect a significant degree of Ca depletion in BP Tau, with the best fit and median Ca abundance of $\rm [Ca/H] = -2.0$ (0.1\% solar). }

In general, the physical parameters obtained from the top-25 distribution are consistent with those of \halpha, with the major difference found in the width of the accretion flow (\rw), \added{and the temperature of the flow (T$_{\rm max}$), with the \CaIIk\ preferring }sightly higher accretion rates and hotter accretion columns with narrower geometries \added{(see Table \ref{tab:lines_results})}. This difference is not fully unexpected as previous modeling of line profiles has found that high transition energies trace more compact flows located within the larger, more diffuse flows traced by \halpha\ \citep{wendeborn_multiwavelength_2024-1}.

\subsubsection{\MgIIk\ line }

The \MgII\ line profiles have complex shapes, due to a combination of blue-shifted absorption features (originating from winds; see Appendix \ref{ap:winds}) and a central absorption feature, which are not included in the magnetospheric accretion model. Therefore, we exclude the regions $-230$ to $-115$ $\rm \kms$, $-95$ to $-55$ $\rm \kms$, and $-30$ to $30$ $\rm \kms$ when fitting the line profile.
As a first method, we tried to model the \MgII\ 2796.4 \AA\ line similarly to \CaIIk, exploring the abundance parameter for the top-100 parameter combinations that can reproduce \halpha. However, this did not result in a good fit of the profile. We attribute this mainly to the fact that the optical and UV data are not contemporaneous, with a 28-yr time difference; therefore, changes in the magnetospheric accretion flow parameters are expected. Hence, we explore the entire parameter space while modeling \MgII\ 2796.4 \AA. 

\begin{figure*}[t!]
\epsscale{0.98}
\plotone{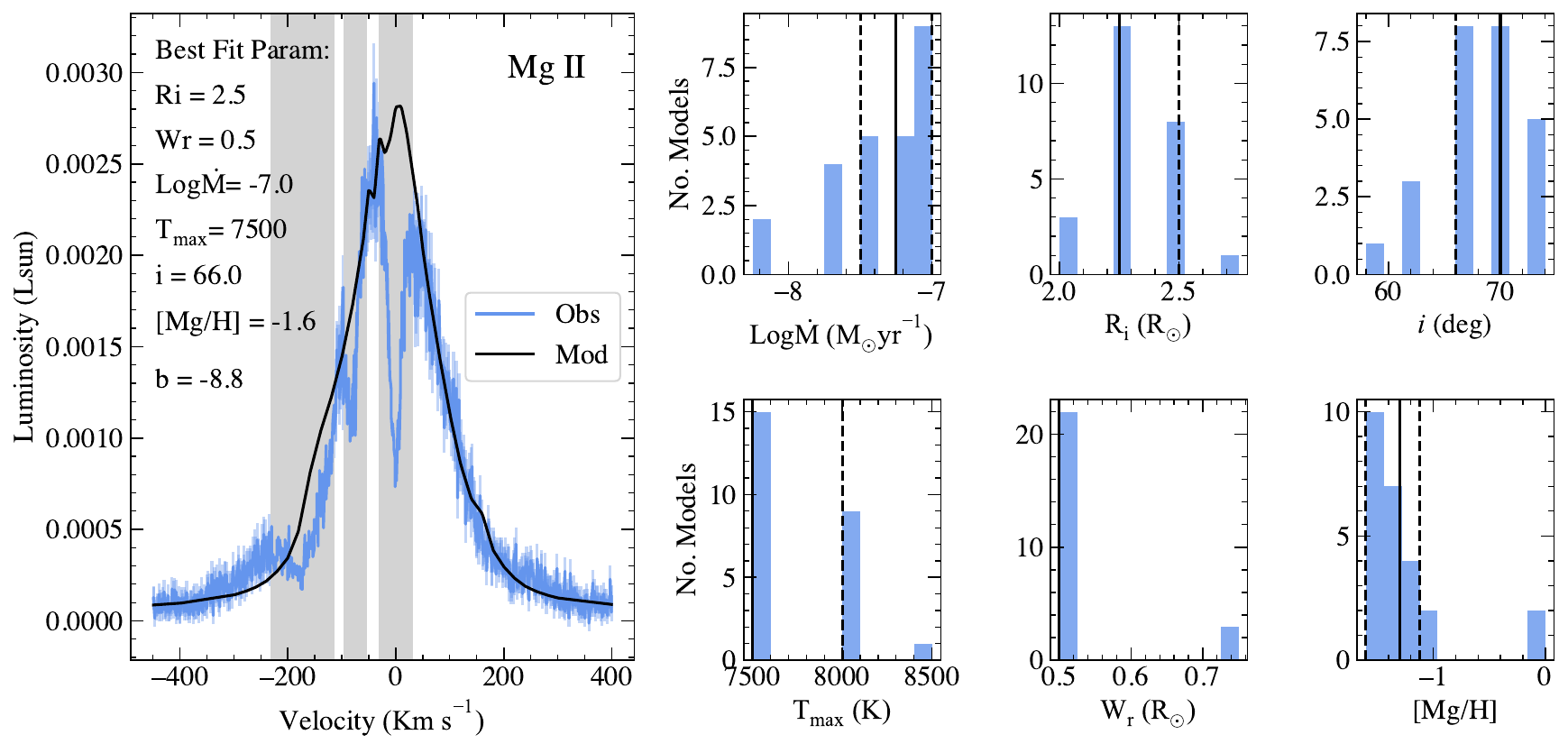}
\caption{Left: Comparison of the best-fit model and the observations for \MgIIk. Grey regions are excluded from the fitting. Right: Parameter distribution for the top 25 best-fitting models. The solid line is the median value for each parameter, dashed lines represent the 25th and 75th percentile. }
\label{fig:mgii}
\end{figure*}

Figure \ref{fig:mgii} shows the profile comparison between the observations and the best-fit model (left) and the parameter distributions for the top 25 best-fitting models (right-hand histograms). \added{The median, 25th and 75th percentiles value of the parameters distributions are shown in Table \ref{tab:lines_results}.} We find that the \MgIIk\ line originates from denser (higher \mdot), colder accretion flows with slightly smaller inner radii (2.50 vs. 3.00-3.50 from \CaIIk\ and \halpha).
We also find a depletion of Mg in the accretion flows. The best fit obtains an abundance of Mg of $\rm [Mg/H] = -1.60$ (2. 5\% solar), and the top-25 distribution a median abundance of $\rm [Mg/H] = -1.30^{+0.20}_{-0.30}$, which translates to a median abundance of 5.0\% of the solar value. The finding of significant degrees of depletion for both Mg and Ca reinforces the idea that depletion is a consequence of a dust-related process in the disk. We further discuss the implications of this result in \S\ref{sec:ca-mg}.

\added{Overall, we find our line profile modeling results (cf. Table \ref{tab:lines_results}) reflect the highly inhomogeneous nature of accretion from the disk, as indicated by numerical simulations \citep{zhu_global_2024, zhu_global_2025}, 
suggesting that \CaIIk\ arises in compact, hot flows, while lines of higher optical depths, such as \MgII\ 2796.4 \AA\ and especially \halpha, trace cooler regions. As mentioned in sections above, this is consistent with previous results of modeling the \hbeta\ and \hgamma\ lines \citep{wendeborn_multiwavelength_2024-1}.
These results, in turn, 
agree with multi-column modeling of the accretion shocks on the stellar surface, which indicates smaller surface filling factors for the denser, higher energy accretion columns \citep{ingleby_accretion_2013,espaillat_measuring_2021,pittman_towards_2022,pittman_odysseus_2025}.}

\begin{deluxetable*}{rcccccccc}[t!]
\tablecaption{Line Profile modeling results\label{tab:lines_results}}
\tablehead{\colhead{} & \colhead{$\rm Log\ \dot{\text{M}}$ ($\msunyr$)} & \colhead{T$_{\rm max}$ (K)} & \colhead{R$_{\rm i}$ (R$_{*}$)} & \colhead{W$_{\rm r}$ (R$_{*}$)} & \colhead{$i$(deg)} & \colhead{ $\rm [X/H]$} & \colhead{ $\rm b$} & \colhead{ $\rm \chi^2_{\nu}$}}
\startdata
    \multicolumn{9}{c}{\halpha}\\
    \hline
    Best fit & $-7.75$  & 8500 & 3.00 & 1.00 & 50 & ... & $-0.7$ & 1.01\\
    Top 25  & $-8.00^{+0.25}_{-0}$  & 9000$^{+500}_{-500}$  & 3.00$^{+0.25}_{-0.25}$ & 1.00$^{+0.25}_{-0.25}$  & 54$^{+0}_{-4}$ & ... & $-0.7$& $1.01-1.36$\\
    \hline
    \multicolumn{9}{c}{\CaIIk}\\
    \hline
    Best fit&  $-7.50$ & 10500  & 3.50  & 0.50  & 78 &  $-2.0$ & $-0.6$ & 1.01\\
    Top 25  & $-7.50^{+0.25}_{-0.25}$  & 10500$^{+0}_{-0}$  & 3.50$^{+0}_{-0.25}$  &  0.50$^{+0.00}_{-0.00}$ & 58$^{+20}_{-4}$  & $-2.0^{+0.1}_{-0.0}$ & $-0.6$ & $1.01-1.48$\\
    \hline
    \multicolumn{9}{c}{\MgII\ 2796.4 \AA}\\
    \hline
    Best fit &  $-7.00$ & 7500  &  2.50 & 0.50 & 66 & $-1.6$ & $-8.8$ & 1.2 \\
    Top 25  &$-7.25^{+0.25}_{-0.25}$  &$7500^{+500}_{-0}$  & $2.25^{+0.25}_{-0.00}$ & $0.50^{+0.00}_{-0.00}$ & $70^{+0}_{-4}$ &$-1.3^{+0.2}_{-0.3}$  & $-8.8$ & $1.2-1.4$ \\
\enddata
\end{deluxetable*}

\begin{figure*}[t!]
\epsscale{0.75}
\plotone{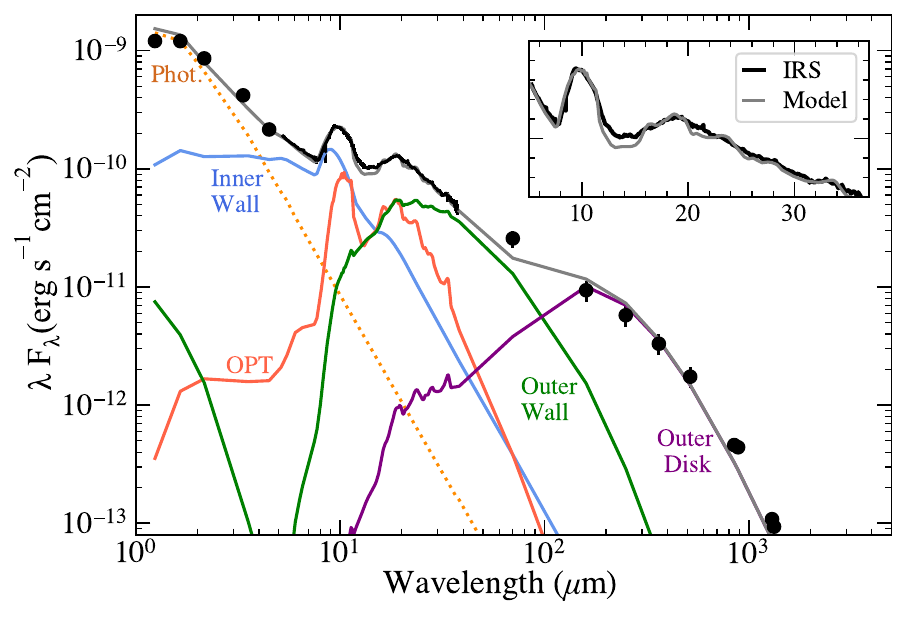}
\caption{Spectral energy distribution of BP Tau. In black, observational data includes photometry and an Spitzer/IRS spectra (references in text), corrected for reddening. The emission from the inner disk wall (blue), the optically thin region \added{(red)}, the outer disk wall (green), and the outer disk (purple) are also shown. The photosphere is shown as a dotted line. An inset of the region of the silicate features is shown in the upper right.}
\label{fig:sed}
\end{figure*}

\begin{figure*}[t!]
\epsscale{0.75}
\plotone{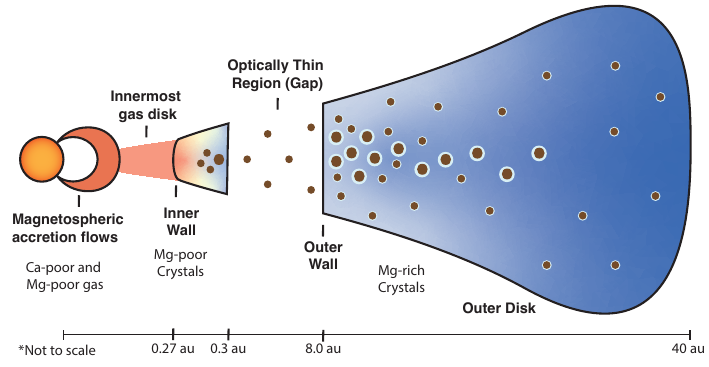}
\caption{Schematic view of the of BP Tau. The composition of the gas in the accretion flows (proxy for the innermost gas disk) is obtained from the emission line profile modeling (see \S \ref{sec:line-mod}). Dust composition for each disk region and their location is obtained from the SED modeling see (\S \ref{sec:sedmodel}). We find a clear change in the Mg content of the crystalline olivine with radius, discussed in \S \ref{sec:dust-discuss}.
}
\label{fig:drawing}
\end{figure*}

\subsection{SED modeling}
\label{sec:sedmodel}

The SED of BP Tau
was fitted by \citet{rilinger_determining_2023} 
using the DIAD model and an artificial neural network \citep{ribas_modeling_2020}, assuming that BP Tau is a full disk. We started our modeling using results from \citet{rilinger_determining_2023}, to obtain initial values for the overall disk structure, such as the degree of settling $\epsilon$ and the viscosity parameter $\alpha$.
However, 
\citet{rilinger_determining_2023} 
adopted a fixed dust composition of
40\% graphites, 30\% olivines and 30\% pyroxenes, and a dust-to-gas mass ratio of 0.01, following
\citet{ribas_modeling_2020}.
In this paper, we focus on the composition of the dust and aim to fit the IRS spectrum in addition to the overall SED.

We could not find a good fit to the details of the silicate 10 $\mu$m and 20 $\mu$m features assuming the same dust throughout the disk. This is frequently found in CTTS; for example, in a detailed study of the SEDs of four CTTS, \citet{mcclure_curved_2013} find that the composition of the dust at the inner edge of the dust disk, as indicated by the wall at the dust sublimation radius, is different from that of the disk, which is representative of regions farther out. In addition, in their study of the dust in CTTS in Taurus,  \citet{sargent_dust_2009} find that the composition of the dust in the two regions which they use to represent the disk is different.

Although originally thought to be
a full and compact disk \citep{long_compact_2019},
recent studies have indicated that
the disk of BP Tau has structure.
\citet{jennings_superresolution_2022} 
identified substructures in a larger fraction of compact disks than previously found \citep{long_compact_2019}
by using a different approach to interpret the 1.3 mm ALMA data, and in particular, they found indications of substructure inside 15 au in BP Tau. 
\citet{kaeufer_analysing_2023} analyzed the SED of BP Tau using a machine learning approach and found that the disk was discontinuous, better described by a two-zone model. \citet{zhang_substructures_2023} found evidence for a ring around 12 au with non-parametric fits to submillimeter visibilities. Finally, \citet{yamaguchi_alma_2024} used a novel 2D super-resolution imaging technique to reanalyze the archival ALMA data and found a central hole and ring around 9.7 au, confirmed by new high-resolution observations by \citet{gasman_minds_2025}.

In agreement with these indications, we found that a better fit to the SED of BP Tau could be obtained with a gapped disk, that is, with an inner disk and an outer disk, both optically thick, and an optically thin region inside the gap. Figure \ref{fig:sed} shows the observed SED of BP Tau and the best-fit SED model obtained, indicating the emission from each disk region. Figure \ref{fig:drawing} shows a schematic view of the disk structure , indicating the position of each region as obtained from the best-fitting SED model and the composition results of the SED and the line profile modeling.

To search for the best fit, we calculated \added{}{separate} models for the inner disk edge (which represents the inner disk emission), for the outer disk, including its inner edge \added{and the proper disk,} and for the optically thin region filling the region between the inner and outer disks. 
The parameters are listed in Table \ref{tab:disk_model} and Table \ref{tab:dust_prop}.
The best fit is searched by minimizing the reduced $\chi^2$ (\S\ref{sec:chi2})
However, changing all the parameters at once became prohibitive. Instead, we searched for the best fit in two steps. Since the emission below $\sim 14 \mu$m comes mostly from the inner wall and the photosphere (cf. Fig. \ref{fig:sed}),
we first searched for a good fit to the 10 $\mu$m feature and the near-IR excess over the photosphere by fitting the spectrum below 14 $\mu$m with a composite model of the photosphere, inner wall and optically thin region. Then, keeping the inner wall model fixed, we searched for the best fit to the overall SED.

\begin{figure}[t!]
\epsscale{1.1}
\plotone{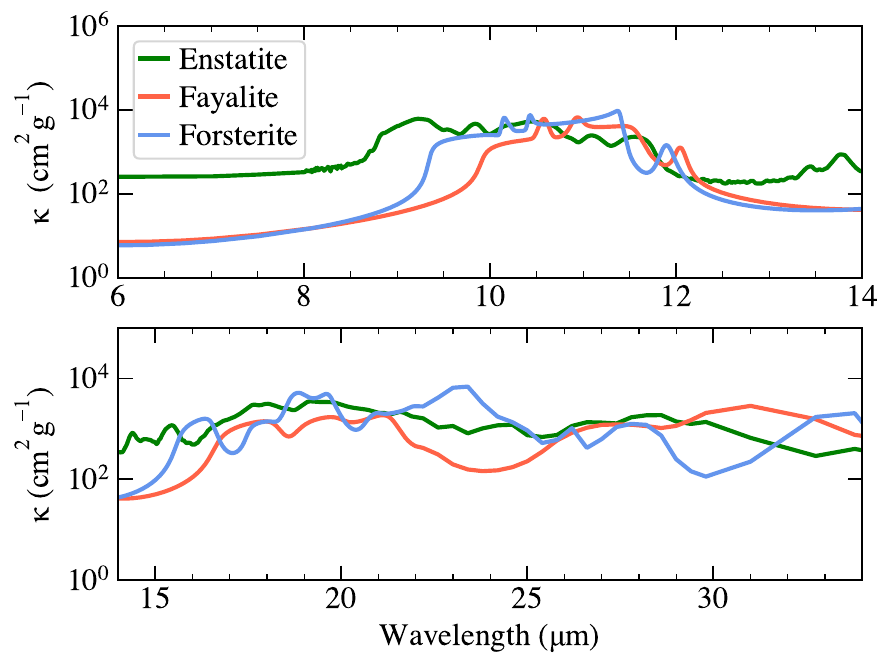}
\caption{Total opacity per gram of dust for crystalline material: enstatite (green), Mg-rich crystalline olivine (forsterite, blue), Fe-rich crystalline olivine (fayalite, red). The maximum size of the grains is a$_{max} = 0.25 \mu$m.}
\label{fig:crystals}
\end{figure}

We aimed to determine the stochiometry and Mg content of the crystalline material in our fitting procedure. Figure \ref{fig:crystals} shows the total opacities for small grains (a$_{max} = 0.25 \mu$m) of enstatite, Mg-rich crystalline olivine (forsterite) and 
Fe-rich crystalline olivine (fayalite).
The peak at $\sim$ 9.2 $\mu$m  and the dip at $\sim$ 32.9 $\mu$m clearly distinguish enstatite from crystalline olivine. Similarly, the short-wavelength extension of the 10 $\mu$m feature, and the shift of the 33 $\mu$m feature peak to longer wavelengths as the Mg content increases help determine the Fe-to-Mg ratio in crystalline olivine. 

The structural parameters of each region are shown in Table \ref{tab:disk_model} and the composition of the dust in each region in Table \ref{tab:dust_prop}. In column 2 of these tables we show the range of values explored in our fitting procedure, in column 3, the results of the top 100 model combination with the lowest $\chi^2$, and in column 4 the adopted best-fit for the parameter. Some parameters are dependent on others, so for those we only give the best fit. We next describe our results for each region.

\begin{deluxetable}{lccc}[t!]
\tablecaption{Disk structural parameters \label{tab:disk_model}}
\tablehead{
\colhead{Parameter} & \colhead{Range} & Top 100 & \colhead{Best fit}  
}
\startdata
\hline
\hline
\multicolumn{4}{c}{Inner wall} \\
\hline 
         $T_{wall,iw}$ (K) &-- & --  & 1400  \\  
         $R_{wall,iw}$(au) & -- & -- & 0.27  \\
         $z_{wall,iw}/H$  & 0.5, 1, 2, 3, 4 & 4$^{+0}_{-0}$ & 4 \\
         $z_{wall,iw}(au)$ & -- & --  & 0.03\\
         $a_{max,iw}$ ($\mu$m) & 0.25, 1 & 0.25$^{+0.00}_{-0.00}$ & 0.25 \\
                  $i$   & -- & -- &   65$^{\circ}$ \\
\hline
 \multicolumn{4}{c}{Outer wall} \\
 \hline
         $T_{wall,ow}$ (K)  & 165, 170 & 170$^{+0}_{-0}$ & 170 \\ 
         $R_{wall,ow}$ (au)  & -- & -- & 8.4  \\
           $z_{wall,ow}/H$   & 0.5, 1, 2, 3, 4 & 0.5$^{+0}_{-0}$ & 0.6 \\ 
          $z_{wall,ow}(au)$   & -- & -- & 0.4 \\ 
         $a_{max,ow}$ ($\mu$m) & 0.25, 1 & 0.66$^{+0.34}_{-0.41}$ &  1.0 \\
                  $i$   & -- & -- &   38$^{\circ}$ \\
 \hline
 \multicolumn{4}{c}{Outer disk} \\
 \hline
        $R_{disk}$ (au) & -- & --  & 40  \\  
        $\alpha$  & 0.0025, 0.036 & -- & 0.0025   \\
      $\epsilon$   & 0.00008, 0.00013& -- & 0.00008  \\
        $a_{max}^s$ ($\mu$m) & 0.25, 1 & -- & 1.00 \\
        $a_{max}^b$ ($\mu$m) & 500, 1000 & -- & 1000 \\
        $M_{disk}$ $(\msun)$ & -- & -- & 0.06 \\
                 $i$   & --  & -- & 38$^{\circ}$ \\
 \hline
 \multicolumn{4}{c}{Optically thin region} \\
\hline
        $R_{min}$ (au) & --  & -- & 0.3  \\  
         $R_{max}$ (au) & 5, 8 & 8$^{+0}_{-0}$ & 8   \\
         Power  & -- & --&   0.1 \\
       log $\tau_{10}^i$  & -3, -2.3, -2., -1.3, -1 & -2$^{+0}_{-0}$ & -2  \\
        $a_{max,opt}$ ($\mu$m) & 0.25, 1& 0.54$^{+0.46}_{-0.30}$ &  0.25 \\
     M$_{dust} (M\earth)$ & -- & --  & 3 $\times 10^{-5}$   \\
\enddata
\end{deluxetable}

\begin{deluxetable}{lccc}[t!]
\tablecaption{Dust properties}
\label{tab:dust_prop}
\tablehead{
\colhead{Parameter} & \colhead{Range} & Top 100 & \colhead{Best fit} 
}
\startdata
\hline
\multicolumn{4}{c}{Inner wall (IW)} \\
\hline
         organic/gas & 0.0025, 0.006  & --    & 0.0025  \\
         H$_2$O ice/gas & --  & --    & 0.0 \\
         Olivine & 0.55 - 1 &  0.79$^{+0.01}_{-0.01}$ &  0.8  \\
         Pyroxene & 0.55 - 1 & 0.0$^{+0.0}_{-0.0}$ & 0.0  \\
         Mg (Pyr.) & 40, 100 &  -- & -- \\
         Forsterite & 0 - 0.5  & 0.2$^{+0.0}_{-0.0}$ & 0.2  \\
         Mg (Fors.) & 0, 100  & 0.0$^{+0.0}_{-0.0}$& 0.0 \\
         Enstatite & 0 - 0.5  & 0.0$^{+0.0}_{-0.0}$ & 0.0  \\
         Silica & 0, 0.005 & 0.01$^{+0.01}_{-0.01}$ & 0.0 \\
         \hline
 \multicolumn{4}{c}{Outer wall (OW)} \\
 \hline
         organic/gas & --  & --    & 0.0025  \\
         H$_2$O ice/gas & 0.00001,0.002,0.004 &   0.002$^{+0.000}_{-0.000}$ & 0.002  \\
         Olivine & 0.5 - 1 & 0.005$^{+0.005}_{-0.005}$  & 0.0  \\
         Pyroxene & 0.5 - 1 & 0.82$^{+0.02}_{-0.02}$ & 0.8  \\
         Mg (Pyr.) & 40, 100 & 100$^{+0}_{-0}$   & 100 \\
         Forsterite & 0 - 0.5  &  0.14$^{+0.1}_{-0.1}$ & 0.2   \\
         Mg (Fors.) & 0, 100  & 88$^{+12}_{-12}$ &  100  \\
         Enstatite & 0 - 0.5  &  0.09$^{+0.06}_{-0.10}$& 0.0  \\
         Silica & --  & --   & -- \\
         \hline
 \multicolumn{4}{c}{Optically thin region (OPT)} \\
\hline
         organic/gas & 0.0025, 0.006  & 0.0029$^{+0.0004}_{-0.0004}$    & 0.0025  \\
         H$_2$O ice/gas & -- & --    & 0.0 \\
         Olivine & 0.5 - 1 &  0.63$^{+0.25}_{-0.12}$ & 0.75  \\
         Pyroxene & 0.5 - 1 & 0.12$^{+0.23}_{-0.12}$ & 0.0  \\
         Mg (Pyr.) & 40, 100 &  60.4$^{+39.6}_{-20.4}$ & -- \\
         Forsterite & 0 - 0.5  & 0.16$^{+0.11}_{-0.04}$ & 0.2   \\
         Mg (Fors.) & 0, 100  & 75$^{+25}_{-0}$ & 100  \\
         Enstatite & 0 - 0.5  & 0.03$^{+0.03}_{-0.03}$& 0.0   \\
         Silica & 0., 0.05  & 0.049$^{+0.001}_{-0.001}$   & 0.05 \\
\enddata
\tablecomments{The dust-to-gas mass ratio of silicates is fixed at 0.004. The Mg fraction of olivine is fixed at 50\% and that of enstatite at 96\%. The abundances of the outer wall and the small grains in the outer disk are assumed to be the same.}
\end{deluxetable}

\subsubsection{Inner disk}

\label{sec:innerdisk}

The inner disk emission is dominated by the frontally irradiated edge of the disk, which we represent by a vertical wall at the dust sublimation radius. In our procedure \citep{dalessio_models_2004}, this is the radius at which the temperature of optically thin dust becomes equal to the sublimation temperature of the silicates, which we kept fixed at 1400K. We solve for the temperature structure in the ``atmosphere'' of the wall, which depends on the stellar and accretion shock emission irradiating the wall, 
and on the sizes and composition of the dust absorbing those radiation fields (Table \ref{tab:dust_prop}). By fitting the emission of this region to the observations, we find that the sublimation radius is 0.27 au $\sim$ 28 $R_*$ (Table \ref{tab:disk_model}), which is larger than the typically assumed $\sim$ 0.1 au. 
We note that we represent the edge of the disk as a single vertical wall; however, the silicate sublimation front is much more complicated. 
The sublimation temperature depends on the
silicate composition
\citep{lodders_solar_2003,lodders_solar_2010} and the gas density
\citep{pollack_composition_1994}; as a result, the edge has a curved shape \citep{isella_shape_2005}. \citet{mcclure_curved_2013} modeled this region with two walls; a lower one with large grains and an upper one with small grains. Since small grains have higher opacities and thus reach the same temperature at larger radii, the upper wall was located at larger radii than the lower one in this study.
The properties of the inner wall in our BP Tau fit (Table \ref{tab:disk_model}) are consistent with those of the upper wall in \citet{mcclure_curved_2013}
in that the grains are small, a$_{max} = 0.25 \mu$m, and the radii are large. As noted in that paper, the emission from a lower wall with larger grains may be too weak to be detected.

The height of the wall acts as a scaling factor that multiplies the SED of the disk model to fit the observations.
We varied this height between 0.5 and 4 scale heights and found the best fit to be 4 scale heights at 0.03 au, consistent with the heights found for other classical T Tauri stars \citep{mcclure_curved_2013}. We assumed that the inclination of the inner disk was consistent with that of the magnetosphere as found from modeling the emission lines (Table \ref{tab:lines_results}, see \S
\ref{sec:misalignment} for a discussion).

To obtain the stoichiometry of the silicates in the inner wall, we varied the fractional abundances of the possible constituents. To scan the parameter space following observed trends \citep{sargent_dust_2009,watson_crystalline_2009}, we assumed that the fractional degree of crystallinity in silicates could vary between 0 and 0.5 of the total silicate dust-to-gas mass ratio, and that the rest would be distributed between olivine, pyroxene, and silica. 
The silica fraction can vary between 0 and 0.05, and the olivine fraction can vary between 0 and 1 of the amorphous material. Among the crystalline material, the forsterite fraction could be 0 to 1, with the enstatite fraction the rest. 
Finally, from the available stoichiometries (\S\ref{sec:optool}), we considered only two values of the Mg fraction in forsterite, 0 and 1, to test extremes while saving computational resources. Similarly, we consider two values in pyroxene, 0.4 and 1. 
We fix the Mg fraction at 0.5, since only 0.4 and 0.5 were available. 

\begin{figure*}[t!]
\epsscale{0.98}
\plotone{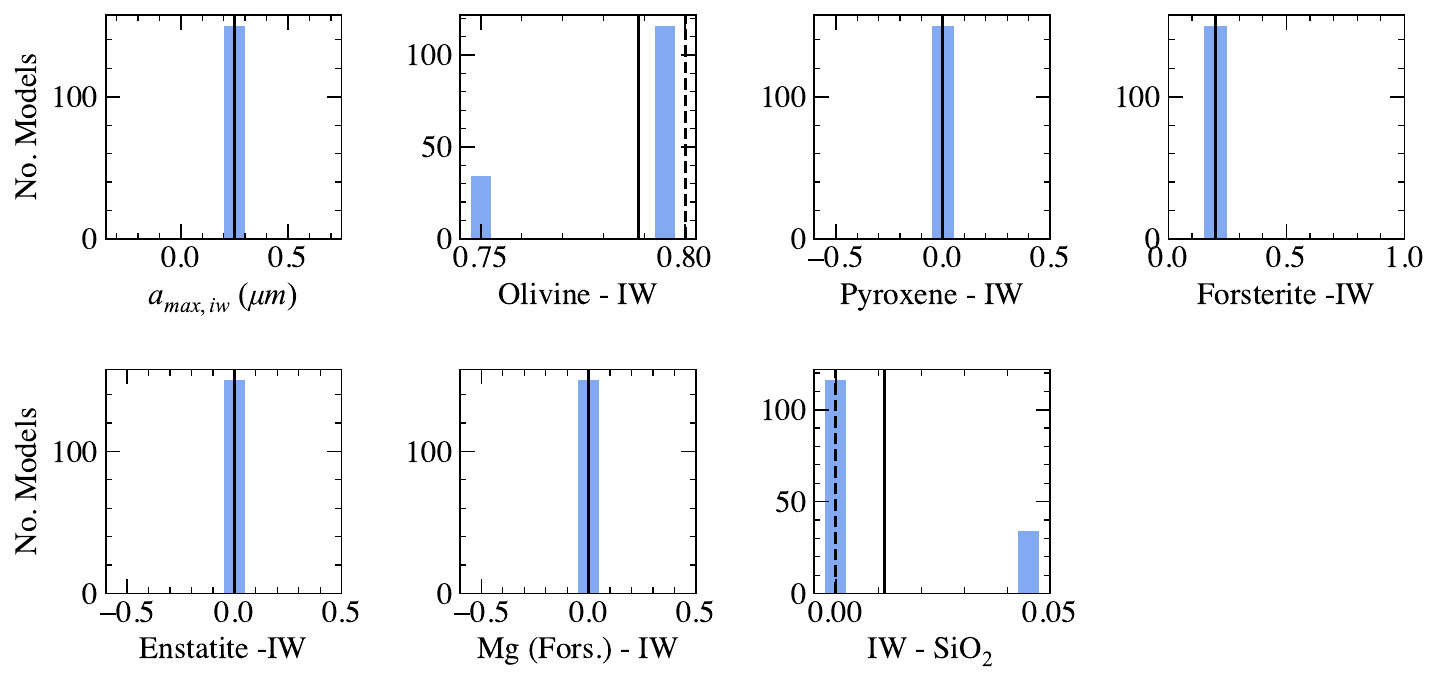}
\caption{Parameter distribution for the top 150 best-fitting models for the inner wall structural parameters and dust properties. The solid line is the median value for each parameter, dashed lines represent the 25th and 75th percentile. }
\label{fig:pariw}
\end{figure*}

In total, we calculated 768 models of the inner wall; the distribution of the physical parameters among the top 150 is shown in Figure \ref{fig:pariw}. The best-fit model indicates that about 80\% of the silicates in the wall are amorphous olivines and 20\% are crystalline fosterite, without pyroxenes or enstatite. In addition, all fosterite has a 100\% Fe content (properly, fayalite). No traces of silica are found. A detailed fit to the SED around the  10 $\mu$m feature is shown in Figure \ref{fig:sed}.

\subsubsection{Outer Wall and disk}
\begin{figure*}[t!]
\epsscale{0.98}
\plotone{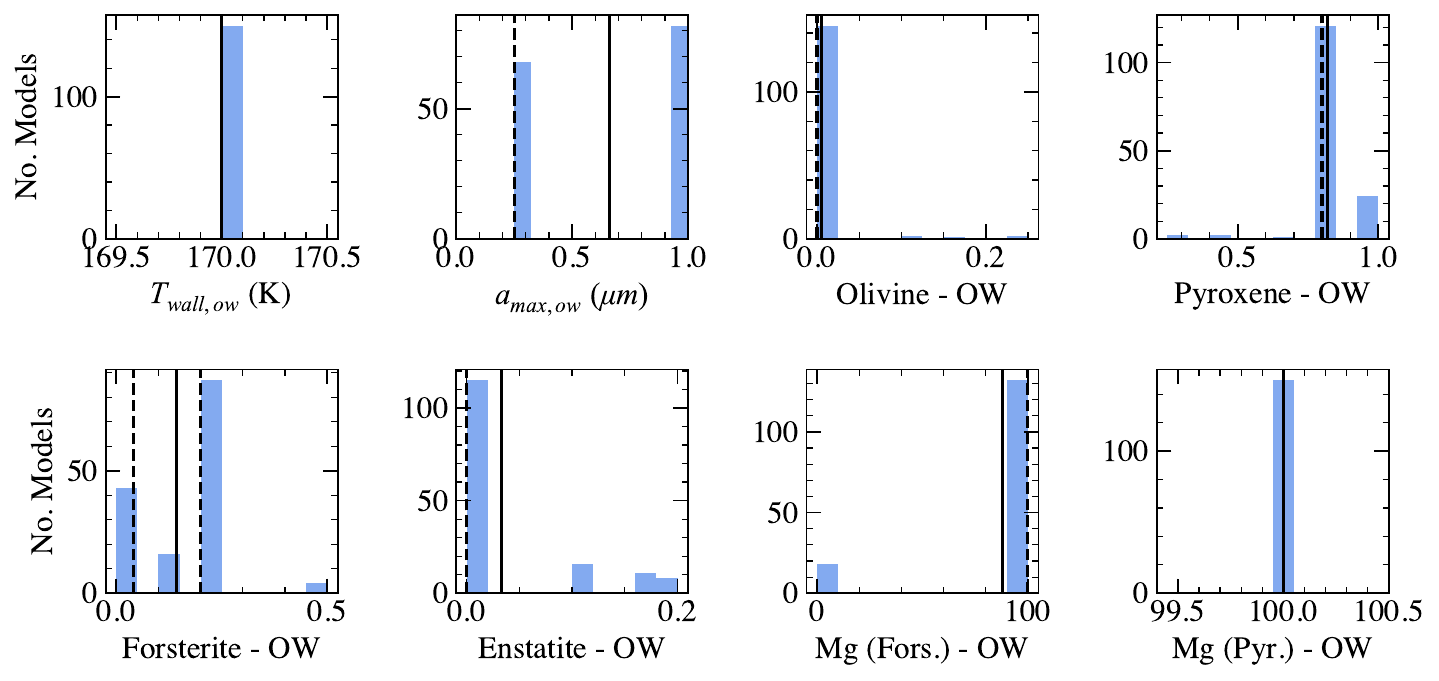}
\caption{Parameter distribution for the top 150 best-fitting models for the outer wall structural parameters and dust properties. The solid line is the median value for each parameter, dashed lines represent the 25th and 75th percentile.}
\label{fig:parow}
\end{figure*}

The outer wall defines the edge of the outer disk,  
assumed to be vertical and frontally illuminated by the star. Similarly to our treatment of the inner wall, we varied the temperature of the top of the wall ``atmosphere'' and calculated the structure and emission of this region, given the properties of the dust.
We run a grid of 2301 models for the outer wall for parameters in the ranges shown in Tables \ref{tab:disk_model} and \ref{tab:dust_prop}. The distribution of parameters for the top 150 models is shown in Figure \ref{fig:parow}. 
Our best fit has a temperature of 170 K, which corresponds to a radius of 8.4 au. This radius is consistent with those determined by reanalyzing archival ALMA data\citep{zhang_substructures_2023,yamaguchi_alma_2024} and new ALMA observations \citep{gasman_minds_2025}.

Given the wall temperature, water ice becomes an important contributor to the opacity. We varied the H$_2$O abundance between essentially no ice and an ice-to-gas mass ratio comparable to that of silicates, and found a best fit for an ice-to-gas mass ratio of 0.002. We found that the amorphous silicates are in pyroxene, in contrast to the inner disk. Similarly, 20\% of the silicates are crystalline forsterite; however, the forsterite is 100\% Mg, in contrast to the inner disk. 

To calculate the structure and emission of the outer disk, we used the stellar and accretion parameters shown in Table \ref{tab:bptau_param} and the structural parameters in Table \ref{tab:disk_model}. The radius and inclination of the disk were taken from \citet{long_compact_2019}. The viscosity and settling parameters were varied within the range found by \citet{rilinger_determining_2023}. We adopted the same dust properties of the outer wall for the small grain population ($a_{max}^s$) and explored the range shown in Table \ref{tab:disk_model} for large grain sizes in the midplane ($a_{max}^b$). We do not perform a large parameter study for the disk, since we are mostly interested in the regions where dust features form, from which we can extract abundance and stoichiometry determinations.

The best fit to the height of the wall is 0.65 scale heights, which corresponds to 0.4 au. To compare this height with that of the disk behind the wall, we take the height above the midplane where the optical depth to the stellar radiation is $\sim 1$ \citep{dalessio_effects_2006}, and obtain $\sim 1$ au. This is higher than the height for the wall that we find; however, 
as in other \added{pre-transitional disks (PTDs)}, it could be explained if the region of the wall closer to the midplane was in the umbra of the inner wall \citep{espaillat_unveiling_2010}, consistent with the possibility of a misaligned inner disk and seesaw variability (\S \ref{sec:midir}.)

\subsection{Optically thin region}

\begin{figure*}[t!]
\epsscale{0.98}
\plotone{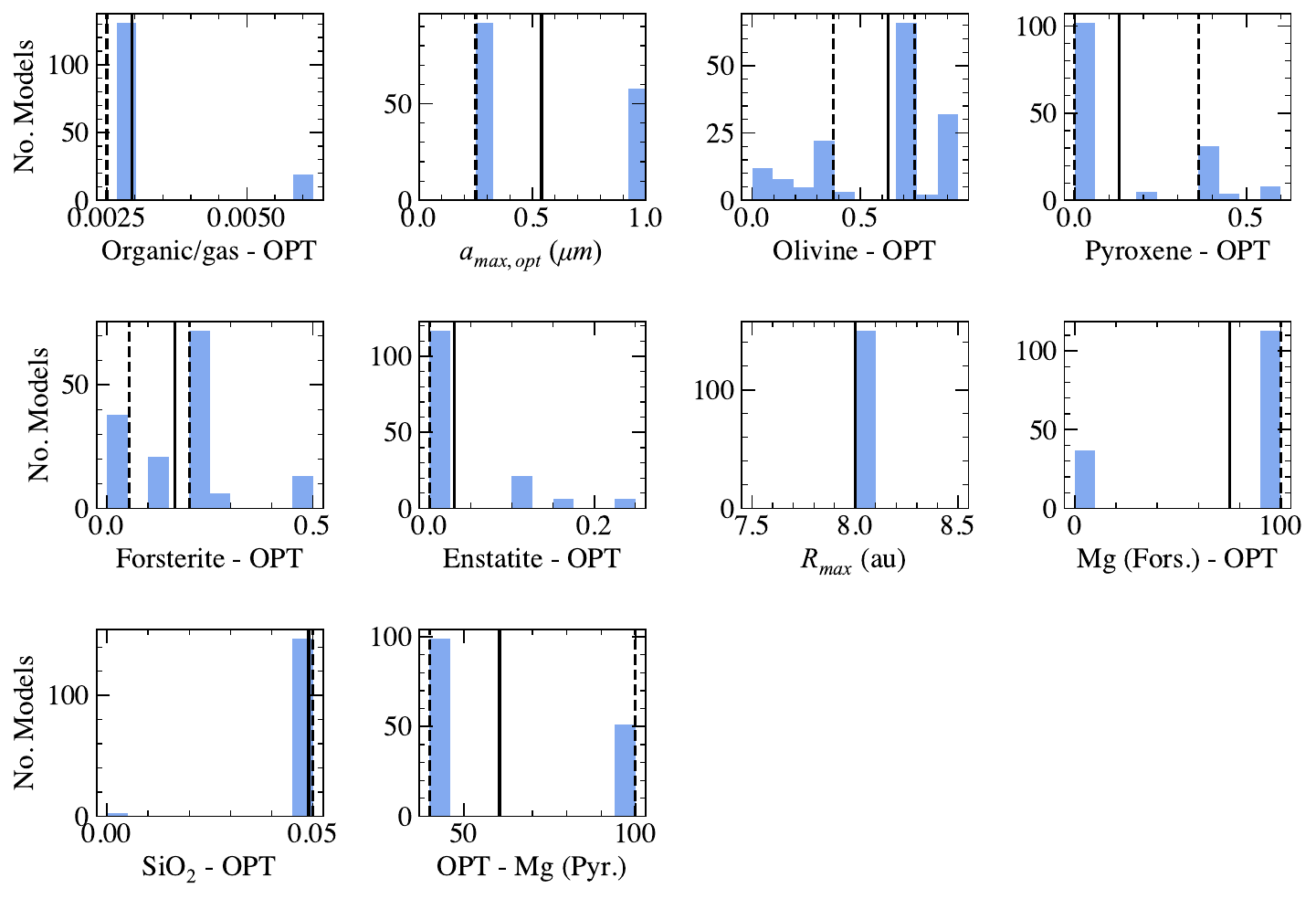}
\caption{Parameter distribution for the top 150 best-fitting models for the optically thin region structural parameters and dust properties. The solid line is the median value for each parameter, dashed lines represent the 25th and 75th percentile. }
\label{fig:paropt}
\end{figure*}

As in other transitional and pre-transitional disks \added{\citep{espaillat_observational_2014}}, we require a region with optically thin dust inside the gap. As in previous studies, we modeled this region using $\tau_{10}^i$, the optical depth at 10 $\mu$m at the inner radius of the region (R$_{min}$) as a parameter, assuming that it varies with radius as
$\tau_{10} = \tau_{10}^i (R_{min}/R)^{{\rm Power}}$. We calculate the temperature at each radius by equating heating to cooling, 
\begin{equation}
\kappa_P^* T_*^4 = \kappa_P T^4
\end{equation}
where $\kappa_P^*$ and $\kappa_P$ are the Planck mean opacities at the stellar and local wavelength range and $T_*$ is the stellar temperature \citep{dalessio_effects_2006}.
The opacities are due to silicates and organics,
and we assume that they are the same in the entire optically thin region. In total, we ran 3072 models; the parameter distribution of the top-150 models and the best fits are shown in Table \ref{tab:dust_prop} and in Figure \ref{fig:paropt}.

The minimum radius of the region is kept fixed at 0.3 au, outside the inner wall. We consider the outer radius R$_{max}$ to be 5 or 8 au, to test the extent of the region. We find that the optically thin material is filling the entire gap. The best fit indicates that the region has a maximum optical depth at 10 $\mu$m of 0.01; the temperature varies from 600K at R$_{min}$ to 150 K at R$_{max}$.
The composition that gives the best fit is shown in Table \ref{tab:dust_prop}. Similarly to the outer wall, the main amorphous material is olivine, but the degree of crystallinity is much higher, 50\%, evenly divided between Mg-rich forsterite and enstatite. Silica also seems to be present. 

We can estimate the dust mass in the optical thin region integrating the dust surface density, scaled to $t^i_{10}$. The dust is assumed to be the same throughout the region, but the carbon sublimates inside 0.6 au, because in this region the temperature is higher than the sublimation temperature for carbonaceous matter. We obtain a mass of 3 $ \times 10^{-5} M_{\earth} \sim $ 0.003 $M_{moon}$, consistent with those found in other transitional and pre-transitional disks \citep{espaillat_probing_2007,espaillat_diversity_2007,espaillat_slowly_2008}.

\section{Discussion} \label{sec:dis}

\subsection{Ca and Mg Depletion in the accretion flows \label{sec:ca-mg}}

\begin{figure}[t!]
\epsscale{1.1}
\plotone{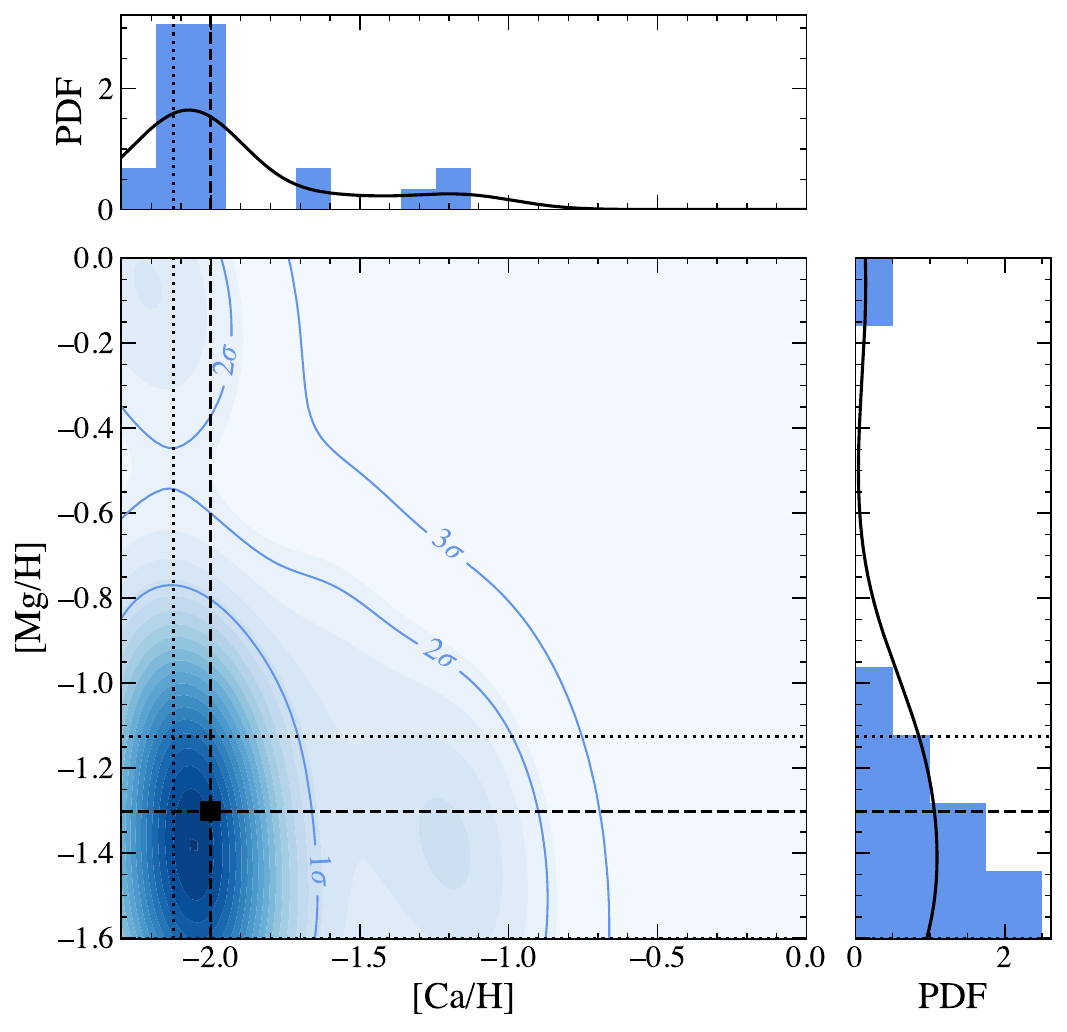}
\caption{Corner plot showing the joint and marginal distributions of $\rm [Ca/H]$ and $\rm [Mg/H]$.Central panel: two-dimensional kernel density estimate (KDE). Darker shades indicate regions of higher probability density. Contours correspond to 
1$\sigma$, 2$\sigma$, and 3$\sigma$ intervals.  
The top and right panels: Probability density distributions of $\rm [Ca/H]$ and $\rm [Mg/H]$, respectively, overlaid KDE curves (black solid lines). Black dashed lines represent the median value of the distribution, and the dotted line show the 25th and 75th percentiles.}
\label{fig:mg-vs-ca}
\end{figure}

The central panel of Figure \ref{fig:mg-vs-ca} shows the 2D kernel density estimate (KDE), which provides a continuous representation of the joint probability density function (PDF) obtained from the best-fitting models for $\rm [Ca/H]$ and $\rm [Mg/H]$. Darker regions indicate areas of higher probability density.
The contours correspond to levels 1$\sigma$, 2$\sigma$, and 3$\sigma$ (68. 27\%, 95. 45\%, and 99. 73\%) of the total probability calculated from the 2D KDE. The top and right panels display the individual PDFs of $\rm [Ca/H]$ and $\rm [Mg/H]$, overlaid KDE curves (solid black lines). In all panels, black dashed lines indicate the median, while dotted lines indicate the 25th and 75th percentiles.

In general, our line profile modeling results indicate that the material in the BP Tau accretion flows is significantly depleted in refractory elements with median abundance values of $\rm [Ca/H] = -2.0^{+0.1}_{-0.0}$ and $\rm [Mg/H] = -1.30^{+0.2}_{-0.3}$. 
This is remarkably subsolar, in strict contrast with the metallicity measured for BP Tau 
\citep[$\rm Fe/H=-0.07\pm0.06$, consistent with solar abundance,][]{dorazi_chemical_2011}
and the measured metallicity for the Taurus-Auriga association, also consistent with solar abundance \citep[][]{santos_chemical_2008, dorazi_chemical_2011}. Therefore, the level of refractory depletion found in BP Tau cannot be explained by initial composition differences from the parent cloud or effects on the emission lines due to stellar activity \citep{spina_how_2020}.

Instead, it must be the consequence of a dust-related process on the disk. Possible causes are the effects of radial drift reducing the
abundance of rocky material that will reach the inner regions of the disk \citep{huhn_how_2023} 
or the trapping of large grains in the pressure bump due to the gap in the disk \citep[e.g.,]{kama_fingerprints_2015,mcclure_carbon_2019,mcclure_measuring_2020,guzman-diaz_relation_2023}. In the case of BP Tau, it is most likely a combination of the two. To properly disentangle the effects of both, it is necessary to have predictions of radial and temporal elemental abundances of refractories from partition models for dust evolution; however, this is beyond the scope of this paper. Moreover, although both Ca and Mg are significantly depleted, the material in the accretion flows appears to be more depleted in Ca than in Mg.

A possibility for the difference in depletion between Ca and Mg is that the composition of the inner gas disk, and therefore the accretion flows, changed in the 28-year gap between the UV and optical observations. However, we do not think this is the case for BP Tau because: (1) It has not gone through a major dust processing event, such as an outburst \citep[e.g.][]{abraham_episodic_2009}, 
since the variability observed in its light curve is attributed to rotational effects and occultation of the hot accretion spot \citep{wendeborn_multiwavelength_2024,burlak_causes_2025}. (2) 28 yrs is shorter than the time it takes for the composition to change given the viscous time scale \added{\citep[> 3 Myr, following eq. (118) of][]{hartmann_accretion_2009}\footnote{\url{https://sites.lsa.umich.edu/lhartm/book/}}}. Therefore, we suggest that the difference is a consequence of the condensation sequence.

In our solar system, the most refractory condensates are Ca-Al-Ti bearing compounds. The major carrier of Ca is Hibonite with a sublimation temperature of 1659 K. Meanwhile, Mg is carried by silicates, such as enstatite and forsterite, with sublimation temperatures of 1316 K - 1354 K, respectively \citep[e.g.,][]{lodders_solar_2003}. 
This difference $\sim 300$ K in the condensation sequence means that in the innermost regions of the disk, Mg sublimates, enriching the gas, while Ca remains in the pebbles, continuing to drift inwards radially faster than the gas. Overtime, this means that the innermost regions will be depleted in Ca sooner than in Mg \citep[e.g.][]{huhn_how_2023}, possibly explaining the difference in the $\rm [Ca/H]$ and $\rm [Mg/H]$ abundances measured
in the accretion flows of BP Tau. Hühn \& Bitsch (private comm.) find that Al, found in species with sublimation temperatures similar to Ca 
depletes faster than Fe, which has lower sublimation temperatures \citep[T $\sim$ 1300,][]{lodders_solar_2003}. 
Calculations of the evolution of dust species in protoplanetary disks for different refractory elements, with parameters characteristic of TTS, are needed to compare with the observed depletion of refractories.

\subsection{Possible inner disk misalignment}\label{sec:misalignment}

High-resolution ALMA imaging of BP Tau estimated a disk inclination between 23.6$^{\circ}$ \citep{gasman_minds_2025} and $38^{\circ}.2^{+0.5}_{-0.5}$ \citep{long_compact_2019}. Our line profile modeling with magnetospheric accretion models yields inclinations consistently higher than the one of the disk ($50 - 70^{\circ}$). The work of \citet{wendeborn_multiwavelength_2024-1} modeling the \halpha\ line profile of BP Tau also obtained a higher inclination than that of the disk, with a median value of $43^{\circ}.7 \pm 1^{\circ}.3$, the highest difference between the inclinations obtained from profile modeling and the expected values \added{in their sample}. 

Both this and \citet{wendeborn_multiwavelength_2024-1} work fitted the line profiles using the magnetospheric accretion model, which assumes that the rotation and magnetic axes are aligned. This is often not the case; these axes tend to be misaligned by up to $\sim 20^{\circ}$ in the T Tauri stars \citep{donati_magnetic_2007, donati_magnetospheric_2008, donati_magnetospheric_2010, donati_large-scale_2011,long_global_2011}. In particular, previous work for BP Tau finds that the field involves a dipole and an octupole component, tilted $20^{\circ}$ and $10^{\circ}$, respectively, with respect to the rotation axis \citep{donati_magnetospheric_2008,long_global_2011}. However, even considering a $20^{\circ}$ misalignment of the rotation and magnetic axes, the inclination values obtained are not consistent with those from ALMA imaging. This, together with the hints of seesaw variability observed between the Spitzer/IRS and the JWST spectra
\citep[cf.][]{espaillat_evidence_2024}, and the azimuthal asymmetry observed in BP Tau polarimetric image \citep{garufi_sphere_2024}, suggests there is a misalignment of the inner disk with respect to the outer disk. 

We note that we assumed an inclination for the outer disk consistent with submillimeter observations when we constructed the SED (\S \ref{sec:sedmodel}). However, to find
a good fit to observations in the 1~to~14 $\mu$m range we had to assume
a higher inclination \added{($65^{\circ}$)} than the one of the outer disk \added{($23.6-38.2^{\circ}$)}. A lower inclination of the inner wall would have required
the scale height to be higher than 4 (the best-fit value, Table \ref{tab:disk_model}), which is not typical of T Tauri stars \citep{mcclure_curved_2013}. Further suggesting a misalignment between the inner and outer disks.

\subsection{Disk dust compositions} \label{sec:dust-discuss}

Although we explored two different values of the dust-to-gas mass ratio of organics, our fitting procedure agreed with the value of 0.0025, similar to the ISM abundances \citep{draine_optical_1984} used in previous disk studies \citep{mcclure_curved_2013}. However, we did find significant changes in the silicate stoichiometry and degree of crystallinity along the disk. 

As discussed in \S \ref{sec:sedmodel},  we found that the amorphous silicates were in olivines in the innermost disk and optically thin region, while they were in pyroxene in the outer wall. 
We also found that crystals were present everywhere, with a degree of crystallinity of $\sim$ 20\% throughout the disk.
We found a clear change in the Mg content of the crystalline olivine with radius, with pure Mg, forsterite, in the outer wall and pure Fe, fayalite, in the inner wall (cf. Fig. \ref{fig:drawing} \added{and Table \ref{tab:dust_prop}}). The pyroxene in the outer wall was also Mg rich. However, we could not determine whether the amorphous silicates became more Fe-rich in the inner disk, because we did not find
pyroxene in the inner wall. No pyroxene was found in the optically thin region in our best-fit model. Nevertheless, the top 150 models distribution shows that although no pyroxene models were preferred, there is a tendency for lower Mg content in pyroxenes in the models that did contain it. Amorphous olivine opacities were available only for materials with a similar Fe and Mg content.

\citet{sargent_dust_2009} analyzed the dust content of BP Tau using a two-temperature disk model, aiming to represent the inner and outer disk. In contrast to our results, they found that the 431K small dust, representative of the inner region, was dominated by amorphous pyroxene, while the amorphous material of the 103K small dust component was approximately 60\% olivine and 40\% pyroxene. They also found enstatite in their warm disk. However, it is difficult to make a direct comparison, since our disk model consists of three regions, each with its own composition and its temperature gradient. In our model, both the inner wall and the optically thin region contribute to the 10 $\mu$m feature, while the optically thin region and the outer wall contribute to the 20 $\mu m$ feature.
In addition, we used different opacity prescriptions that include the Mg/Fe ratio. We note that \citet{sargent_dust_2009} consider BP Tau to be of ``mixed composition'', with no clear dominant dust component.

We can compare with trends found in the literature for the different types of disks around T Tauri stars.
\citet{mcclure_curved_2013} analyzed in detail the SEDs of four disks taken to be full
with emphasis on the inner wall. They aimed to mimic the complex shape of the dust destruction radius using two walls, a lower one with large grains and an upper one with small grains. As mentioned in \S \ref{sec:innerdisk}, our best fit properties for the inner wall are consistent with the upper wall of \citet{mcclure_curved_2013}, which contributes the most to the silicate feature. This upper wall is composed of
olivines, in most cases, and forsterite. The dust in the outer disks of this sample, modeled separately from the wall, was pyroxene and forsterite with very low abundance, which is consistent with our results for the outer wall and disk of BP Tau. Note that recent observations have detected the structure in at least two disks of this sample \citep{long_compact_2019}.

\citet{olofsson_c2d_2010} analyzed a large sample of disks around Herbig Ae/Be stars and T Tauri stars using a two-temperature model. They did not discuss the stoichiometry of the amorphous silicates, but showed the contribution to the silicate features of enstatite and forsterite of different sizes. They found sub-micrometer grains in their warm regions, which is consistent with those we found for the inner wall. Generally, these small grains have a higher fraction of forsterite than enstatite, with enstatite absent in 50\% of the cases, which is consistent with our findings for BP Tau. Similarly, \citet{jang_dust_2024} only finds forsterite in the inner disk of PDS70.

\citet{espaillat_spitzer_2011} modeled the SEDs of a sample of four transitional disks (TDs) and nine pre-transitional disks. They used a modeling framework similar to ours and considered an inner disk, represented by the wall at the dust destruction radius, an outer disk with its frontally illuminated wall, and an optically thin region between the inner and outer disks.
They modeled the silicate stoichiometry in the outer disk wall and in the optically thin region, when present. They considered olivines and pyroxenes, as well as forsterite, enstatite, and silica. However, they used different prescriptions for the opacities, which did not include the Mg content.

In the work of \citet{espaillat_spitzer_2011}, the bulk of silicates are amorphous but crystals are present in most cases. In addition, they found differences between the PTDs and TDs. In the outer walls of TDs, pyroxene is common and sometimes dominant, while crystalline silicates tend to be forsterite. In contrast, in the outer walls of PTDs, olivine is more common and enstatite can be present along forsterite. 
However, in the \citet{espaillat_transitional_2012} study, the composition of the silicates in the optically thin regions is not clearly different between PTDs and TDs. Olivine is most frequently found; crystals are forsterite and enstatite when found, with fractions $\le$ 20 \% (except for one star), similarly to the outer wall. 

Interestingly enough, the results for disk in BP Tau indicate that it is much closer to a transitional disk than to a pre-transitional disk, even though an inner disk is clearly present. 
This may be an indication of a more advanced evolutionary state in BP Tau than in the typical PTD, consistent with the high degree of depletion found in Ca and Mg, and also with the fact that the timescales for evolution, which are likely proportional to the viscous timescale, $t_v = R_d^2/\nu$, with $\nu$ the viscosity, may be shorter for BP Tau given the small size of its disk (Table \ref{tab:disk_model}).
A detailed study of TDs and PTDs using the same opacities and fitting methods is necessary to provide more firm conclusions.

\citet{bouwman_formation_2008} studied 6 disks in old regions, with ages $>$ 5 Myr, an age range in which more than 90\% of the disks have already dissipated
\citep{hernandez_spitzer_2007}.  
Using a two-temperature model, they found pyroxene and forsterite in the outer disk, similarly to those of BP Tau, but they also found a larger ratio of enstatite to forsterite in the inner disk. These similarities suggest BP Tau is in a more evolved stage than a typical pre-transitional disk.

Overall, BP Tau has proven a rather interesting test source, opening a window for connecting depletion of refractories in accretion flows with direct changes in the dust composition trough the disk and the possible impact of the inner disk structure on the composition of rocky material available to form planets in the inner 10au. However, further work studying a sample of protoplanetary disks is needed to draw conclusions about the implications for planetesimal and planet formation.

\section{Summary and Conclusions} \label{sec:con}

In this work, we aim to characterize the dust composition throughout the disk of BP Tau, as a test study to trace dust evolution and refractory depletion in the innermost regions as evidence for radial drift and dust-trapping in protoplanetary disks. We used the magnetospheric accretion model to analyze the \CaIIk\ and \MgII\ 2796.4 \AA\ emission lines and determine abundances relative to hydrogen for Ca and Mg. We used the D'Alessio irradiated accretion disk models to model the SED and obtain the dust stoichiometry of the 10$\mu$m and 20$\mu$m silicate features. Here we summarize the main results:

\begin{enumerate}
    \item We find a significant degree of depletion of refractory material in the innermost gas disk with best-fit values of $\rm [Ca/H] = -2.0$ and $\rm [Mg/H] = -1.6$, and median abundances of $\rm [Ca/H] = -2.0^{+0.1}_{-0.0}$ and $\rm [Mg/H] = -1.30^{+0.2}_{-0.3}$.
    
    \item Given that both the abundances of Ca and Mg are highly subsolar, in remarkable contrast to the known metallicity of BP Tau \citep{dorazi_chemical_2011}, we attribute this to dust-related
    processes on the disk, such as radial drift and dust trapping due to the known pressure bump/gap in the disk.

    \added{\item 
    Our results suggest that 
    \CaIIk\ arises in compact, hot flows, while lines of higher optical depth, \MgIIk\ and especially \halpha, trace more extended, cooler regions. The difference in 
    geometrical properties derived from the analysis of different emission lines reflects the highly inhomogeneous nature of the accretion flows, as expected from numerical simulations \citep{zhu_global_2024,zhu_global_2025}. This is consistent 
    with previous 
 modeling of line profiles and accretion shocks onto the stellar surface \citep{ingleby_accretion_2013,espaillat_measuring_2021,pittman_towards_2022,wendeborn_multiwavelength_2024-1,pittman_odysseus_2025}.}
    
    \item Our SED modeling recovers a gap extending up to 8 AU, filled with optically thin dust, consistent with the results of submillimeter observations of BP Tau \citep{yamaguchi_alma_2024,gasman_minds_2025}.

    \item We find evidence for a misalignment between the inner and outer disk in BP Tau. Our modeling of emission line profiles and the SED $10\mu m$ feature yields an inclination for the inner disk that is more than $20^{\circ}$ higher than the inclination of the outer disk derived from sub-mm imaging. This exceeds the maximum possible difference between the stellar rotation and magnetic axes found for BP Tau \citep[$\sim20^{\circ}$,][]{donati_magnetospheric_2008}, suggesting the misalignment.
    
    \item We found that the silicates in the outer wall and disk were pyroxene and forsterite, while they were olivines and forsterite in the inner wall. We found a significant decrease of the Mg-to-Fe ratio with decreasing radius, with Mg-rich silicates in the outer wall and pure Fe forsterite (i.e. fayalite) in the inner wall. This is consistent with the Mg depletion we measure in the innermost disk.

    \item BP Tau's dust radial distribution and stoichiometry changes seem consistent with those of transitional disks found in the literature. However, a larger study of the dust stoichiometry of structured disks is needed to provide further connections between structure, dust composition, and dust evolution.

\end{enumerate}

\begin{acknowledgments}
   This work was partially supported by NASA/FINESST 80NSSC24K1550, HST-AR-17047.001-A, and NASA XRP 80NSSC2K0151 grants. \added{E.M.M. acknowledges the support from a  SECIHTI (formerly CONAHCyT) grant.
   }

   We thank Katya Gozman and Lee Hartmann for their meaningful comments. We also thank the HST helpdesk. This paper has benefited from multiple discussions in the MODELA research group.

   The authors also acknowledge and thank the referee for their useful comments, which improved the original manuscript.
\end{acknowledgments}

\facilities{HST, Spitzer, JWST, Las Campanas Observatory}
\software{\texttt{Astropy} \citep[][]{astropy_collaboration_astropy_2013,astropy_collaboration_astropy_2018}, \texttt{PyAstronomy} \citep[][]{czesla_pya_2019},  \texttt{Eniric} \citep[][]{neal_jason-nealeniric_2019},
\texttt{Scipy} \citep[][]{virtanen_scipy_2020}, CarPy \citep{kelson_evolution_2000,kelson_optimal_2003}
}

\appendix

\section{Detailed Fitting of the line profiles \label{ap:fitting-lines}}

In order to identify the best-fitting model, we implemented a maximum likelihood approach. To account for the possible underestimation of observational uncertainties, we introduce an error inflation parameter $b$ that modifies the total variance at each data point $i$: $s_i^2=\sigma_i^2+10^b$, where $\sigma_i$ represents the original uncertainty of the observations \citep[e.g.][]{hogg_data_2010,line_uniform_2015}. This parameter reflects the degree of error adjustment necessary to explain the observed data. The log-likelihood function used to evaluate the fit of a model $F(x)$ to the observed data $y$ is given by:

\begin{equation}
\ln \mathcal{L}(y \mid x)=-\frac{1}{2} \sum_{i=1}^n\left[\frac{\left(y_i-F_i(x)\right)^2}{s_i^2}+\ln \left(2 \pi s_i^2\right)\right]
\label{eq:lk}
\end{equation}

where $n$ is the number of data points.
The first term inside the summation is the $\chi^2$, which penalizes large residuals. The second term is the Gaussian normalization factor; because we include error inflation, the normalization factor can change and has to be taken into account. The inclusion of the second term in the likelihood term prevents the error inflation parameter from approaching infinity.

Given the discrete nature of our forward modeling, the optimization was performed iteratively using the scipy.optimize.minimize Python package:

\begin{enumerate}
\itemsep0em 
    \item For our initial guess of model, we optimize $b$ by maximizing the likelihood.
    \item Using the optimized $b$, the log-likelihood values are calculated for all models on the grid.
    \item The model that maximizes the likelihood is selected.
    \item Steps 1-3 are repeated until the change in log-likelihood between iterations falls below a predefined tolerance threshold, chosen to be $10^{-6}$.
\end{enumerate}

Upon convergence, the final optimal values of $b$ and the corresponding best-fit model parameters are identified. For the purposes of this work, the error inflation parameter is allowed to vary between $0.01 \times \min \left(\sigma_i^2\right)$ and $100 \times \max \left(\sigma_i^2\right)$. Overall,
this method ensures a systematic exploration of the model grid while accounting for potential limitations in measurement uncertainties or model assumptions in a self-consistent manner.

For each line, we present the best fit and the top 25 best-fitting models ranked according to their log-likelihood values after the iterative optimization process. From the top-25 best-fitting models we calculate the median for the parameters and take the 25th and 75th percentiles as the uncertainties. \added{The results of the modeling are shown in Table \ref{tab:lines_results} and discussed in Section \S \ref{sec:line-mod}.}

\section{Outflows detected in the \MgII\ 2796.4 \AA\ line profile }\label{ap:winds}

The profile of the \MgII\ 2796.4 \AA\ line exhibits two blue-shifted absorption features, at $-173.20 \kms$ and $-84.7 \kms$, attributed to mass ejection, that is, winds \citep[e.g.][]{hartmann_accretion_2009}. The high-velocity wind is consistent with the velocity detected one of the Herbig Haro objects associated with BP Tau \citep[Knot A, $v = -160 \pm 5 \kms$,][]{dodin_jet_2024}.
BP Tau also shows a micro (counter-) jet at lower velocities \citep[$v_{mj} = 118 \pm 5 \kms$,][]{dodin_jet_2024}; however, this is inconsistent with the low velocity wind we find. Outflow variability over time can explain why we find a lower velocity jet that is not detected in outflow analysis with recent observations ($\Delta t \sim 31yr$). \\

Following \citet{calvet_properties_1997}, we can estimate the mass loss rate of both winds. For each feature, the optical depth at a given velocity $\rm v$ is given by

\begin{equation}
\tau=\frac{\pi e^2}{m_e c} \frac{f\ c}{\nu_0} \frac{n_l(\mathrm{v})}{d\mathrm{v} / dz}
\end{equation}

Where $\nu_0$ is the line frequency, $f$ is the oscillator strength, $n_l$ is the density of the lower level of the transition of atoms moving at velocity $\mathrm{v}$, and $d \mathrm{v} / d z$ is the velocity gradient ($d\mathrm{v} / dz \sim \mathrm{v_{\infty}} / R_{*}$). The mass-loss rate is given by

\begin{equation}
\dot{M}_w \sim \Delta A\ \mathrm{v}\ \mu\ m_{\mathrm{H}}\ n_{\mathrm{H}}(\mathrm{v}) \sim \Delta A\ \mathrm{v}\ \mu\ m_{\mathrm{H}}\ \eta\ n_l(\mathrm{v}),
\end{equation}

where $\Delta A$ is the cross-sectional area of the wind at $v$ ($\Delta A \sim \pi\ 4 R_{\star}^2$), $\mu$ is the mean molecular weight, $n_{\mathrm{H}}$ is the number density of hydrogen and $\eta \equiv n_{\mathrm{H}} / n_l$. We estimate the optical depth ($\tau$) from the depth of the feature relative to the best-fitting model. In addition, we assume and $\eta$ as the inverse abundance of Mg obtained from the emission line analysis -- that is, $\rm \eta \equiv n_{\mathrm{H}} / n_l \sim 1 / (N_{Mg}/N_{H})$, which assumes that all Mg is at the ground level --. With $\mu = 2.4$ and $\rm [Mg/H] = -1.6$ (best-fit model Mg abundance, relative to solar), we obtain mass loss rates of $\rm Log\ \Mdot_w (-173.20 \kms) = -8.6\ \msunyr$ and $\rm Log\ \Mdot_w (-84.7 \kms) = -9.7\ \msunyr$. This corresponds to $\sim 8\%$ and $\sim 0.07\%$ of the accretion rate from the literature ($\rm Log \Mdot = -7.5 \msunyr$), and $\sim 2\%$ and $0.02\%$ of the accretion rate recovered by our best-fitting model ($\rm Log \Mdot = -7.0 \msunyr$), respectively. Given the optically thick nature of the \MgII\ 2796.4 \AA\ line, these values should only be considered as lower limits on the values of the mass loss rate.

\bibliography{references}{}
\bibliographystyle{aasjournalv7}

\end{document}